\documentclass{egpubl}

\usepackage{sca2020}

\SpecialIssuePaper         % uncomment for final version of Computer Graphics Forum, special issue
\usepackage[T1]{fontenc}
\usepackage{dfadobe}  

\usepackage{cite}  % comment out for biblatex with backend=biber
\BibtexOrBiblatex
\electronicVersion
\PrintedOrElectronic
\ifpdf \usepackage[pdftex]{graphicx} \pdfcompresslevel=9
\else \usepackage[dvips]{graphicx} \fi

\usepackage{egweblnk} 

\usepackage{cite}
\usepackage{amssymb}
\usepackage{amsmath}
\usepackage[percent]{overpic}
\usepackage{csquotes}

\usepackage{appendix}
\usepackage{booktabs} % For \toprule, \midrule and \bottomrule

\newcommand{\norm}[1]{\left\lVert#1\right\rVert}

\usepackage[labelfont=bf,textfont=it]{caption}
\usepackage{subfig}
\title{Efficient 2D Simulation on Moving 3D Surfaces}

\author[Morgenroth et al.]
{\parbox{\textwidth}{\centering 
        D. Morgenroth$^{1}$\orcid{0000-0002-4152-755X},  
        S.Reinhardt$^{1,2}$, 
        D. Weiskopf$^2$\orcid{0000-0003-1174-1026}, 
        and B. Eberhardt$^1$\orcid{0000-0001-6428-4610}
        }
        \\
{\parbox{\textwidth}{\centering 	$^1$Media University Stuttgart, Germany\\
	$^2$Visualization Research Center, University of Stuttgart, Germany
       }
}
}
\begin{document}
	
	% uncomment for using teaser
	 \teaser{
	    \begin{overpic}[width=\linewidth%,grid,tics=10
	    ]{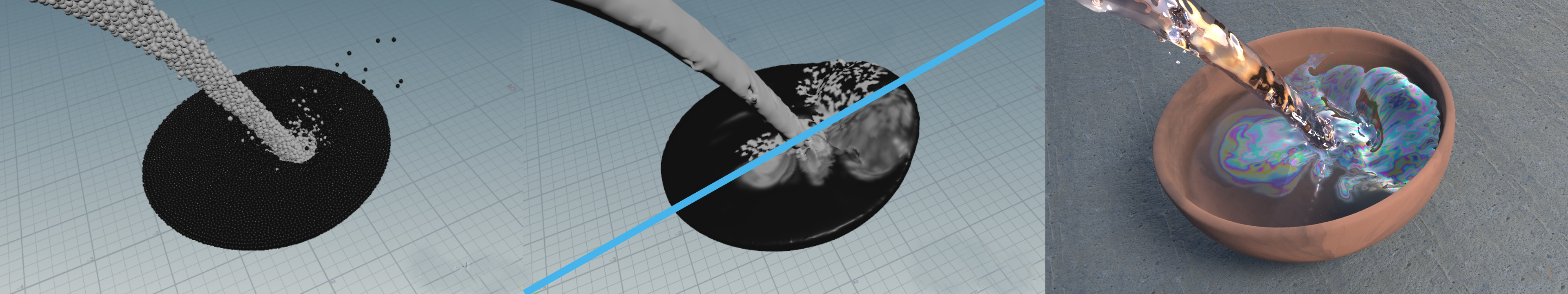}
            \put (2,16) {\color{white} Input Simulation}
            \put (35,16){\color{white} Scalar Field}
            \put (50,2) {\color{white} Surface Simulation}
            \put (90,2) {\color{white} Rendering}
        \end{overpic}
	  \centering
		\caption{Water polluted with oil is poured into a cup. We simulate the oil film on top of the existing fluid simulation. The velocity and mass are taken from the existing animation. On the left is the coarse input simulation. In the upper half of the middle image is the resulting scalar field. The lower half shows the result of our method. Our high-resolution 2D simulation adds convincing visual details to the coarse input simulation. The right image displays the final rendering that needs the fine details from our surface simulation to generate the high-resolution thin-film interference effects.}
	 \label{fig:teaser}
	}
	
	\maketitle
	%-------------------------------------------------------------------------
	\begin{abstract}
	    We present a method to simulate fluid flow on evolving surfaces, e.g., an oil film on a water surface. Given an animated surface (e.g., extracted from a particle-based fluid simulation) in three-dimensional space, we add a second simulation on this base animation. In general, we solve a partial differential equation (PDE) on a level set surface obtained from the animated input surface. The properties of the input surface are transferred to a sparse volume data structure that is then used for the simulation. We introduce one-way coupling strategies from input properties to our simulation and we add conservation of mass and momentum to existing methods that solve a PDE in a narrow-band using the Closest Point Method. In this way, we efficiently compute high-resolution 2D simulations on coarse input surfaces. Our approach helps visual effects creators easily integrate a workflow to simulate material flow on evolving surfaces into their existing production pipeline.
		%-------------------------------------------------------------------------
		%  ACM CCS 1998
		%  (see http://www.acm.org/about/class/1998)
		% \begin{classification} % according to http:http://www.acm.org/about/class/1998
		% \CCScat{Computer Graphics}{I.3.3}{Picture/Image Generation}{Line and curve generation}
		% \end{classification}
		%-------------------------------------------------------------------------
		%  ACM CCS 2012
		
		%The tool at \url{http://dl.acm.org/ccs.cfm} can be used to generate
		% CCS codes.
		%Example:
		

	\end{abstract}	
	%-------------------------------------------------------------------------
	\section{Introduction}
	\begin{figure*}[t]
	 \centering
		\includegraphics[width=0.94\textwidth]{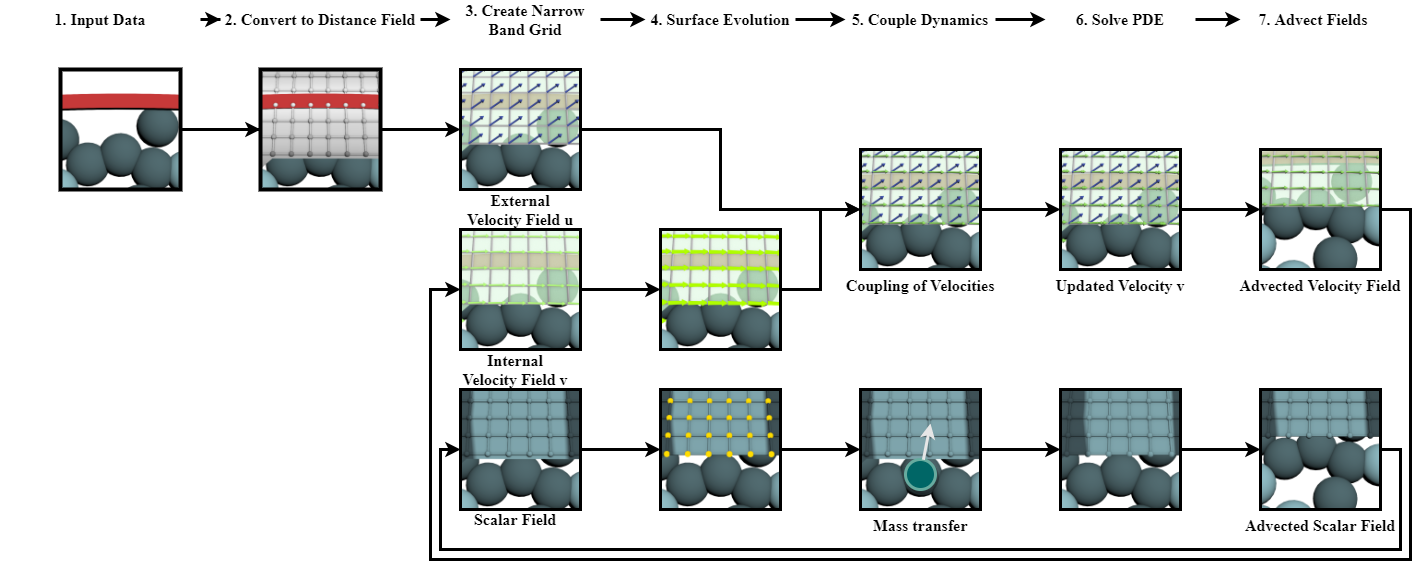}
		\caption{
		The seven steps of our method as a flow chart. 
		The steps are  placed in order of execution from left to right. 
		%The third step creates 3 different grids shown underneath.
		After the input of the external data (Step 1), the data is transferred to the narrow-band (Step 2 and 3).
		Next, the surface is evolved (Step 4) and the outer process is coupled with the inner dynamics (Step 5), before solving the set of PDEs on the surface (Step 6).
		Finally, the fields are advected (Step 7).
		\vspace*{-5pt}
		}
		\label{fig:schematic}
	\end{figure*}
    In typical workflows for generating digital visual effects, a team of VFX artists iteratively refines a given sequence until they achieve good quality. Going from rough storyboards over blocked animation to the final shot with many layers of physical simulations, each shot passes through the VFX pipeline, where domain specialists add new effects and details. With physical simulations, often an effect is finished and approved before secondary effects are layered on top of it. For example, after simulating the fluid flow of a water surface, secondary effects like splashes and foam \cite{takahashi2003realistic} \cite{ihmsen2012unified} are added on top of this \enquote{basic} water simulation, which we will call \enquote{base animation}. In this context, we propose a method to add secondary effects on top of the base animation by solving a PDE on the surface. As an example, we simulate a thin-film 2D fluid simulation on top of a possibly precomputed, bulk fluid simulation as shown in Fig.~\ref{fig:teaser}. We employ a one-way coupling to transfer momentum and mass from the 3D fluid simulation to the surface simulation. We base our coupling on physical derivations and provide them with plausible parameters to control its effects. This coupling allows us to iterate on the secondary effects with a consistent high-resolution 2D simulation on top of the unchanged coarse 3D input simulation.

    The  goal is to solve a partial differential equation (PDE) on a moving 2-manifold. The approach can be used to address different types of problems  that require solving PDEs like fluid flow, reaction-diffusion texture synthesis, or 2D wave equations.
    Fig.~\ref{fig:schematic} depicts the steps of our approach. We start with a moving surface as input geometry. After converting the input data into a distance field and transferring values into a narrow-band grid around the surface, we introduce quantities from the input 3D simulation into our surface domain in a coupling step where the strength of the coupling can be driven by parameters. Then, we use the Closest Point Method (CPM) to embed the 2-manifold in the 3D space of the moving surface and solve 2D PDEs in a 3D narrow-band.

	With our method, we can model secondary effects very efficiently, and we evaluate our approach by simulating a variety of effects such as pouring an oil film into water, simulating reaction-diffusion on a water surface, or the surfactants in a heated soap bubble. 
	%We make our code available as an Open Source project \cite{Cappucino:2020} to simplify the integration into existing production pipelines.  
	The source code is available on GitHub~\cite{Cappucino:2020}. 
	
	\section{Related Work}
	Secondary effects can be simulated on top of the base animation in two different ways. The first way is to directly integrate them into the simulation, for example, bubbles in foam   \cite{thuerey2007real}, \cite{cleary2007bubbling}, or\cite{ihmsen2011animation}. Here, the secondary effects are two-way coupled with the main fluid. Modeling the secondary effects in such a way does not allow for post-processing of existing simulations and animations without the need to re-simulate. To address this problem, we decouple the surface simulation from the input simulation. Furthermore, this decoupling allows us to use a higher resolution on the surface simulation and add fine details on top of coarse inputs.
	
    The second way to add additional effects like splashes, bubbles, and foam is to simulate them with the main fluid but not to exert forces back onto the main simulation, e.g., \cite{takahashi2003realistic}, \cite{kim2012controlling}, or \cite{ihmsen2012unified}. Splashes and bubbles need true 3D solvers as they extend into, or out of, the surface. Foam only lives on the surface and can be rendered \cite{akinci2013screen} by tagging surface particles of the original simulation or by moving texture patches with the surface \cite{gagnon2016dynamic}. In contrast, we are interested in effects that are limited to the 2D surface, but we want to simulate a finer resolution on the surface than the original 3D fluid simulation offers. 
	
	 Solving PDEs on 2-manifolds has been addressed earlier. Some solutions are designed for special cases of geometry, e.g., spheres \cite{hill2016efficient}. Instead, we are interested in a generic solution. Stam~\cite{stam2003flows} simulated fluid flows on static Catmull-Clark surfaces or Shi et al.~\cite{shi2004inviscid} and Vantzos et al.~\cite{azencot2015functional} on static triangle meshes. Azencot et al.~\cite{azencot2014functional} model fluids on triangle surfaces using their vorticity by a time-varying scalar function. Ruuth et al.~\cite{ruuth2008simple} proposed the Closest Point Method (CPM) to calculate PDEs on surfaces. This was applied by Macdonald et al.~\cite{macdonald2013simple} to calculate reaction-diffusion terms on surfaces derived from point clouds. CPM on static surfaces can be calculated in real time \cite{auer2012real}. The theory for solving PDEs on evolving surfaces is explained by Xu and Zhao~\cite{xu2003eulerian}. Examples are the simulation of soap film on bubbles \cite{isnhw2020soapfilm_with_thickness} where the PDE is solved on a triangle  mesh, or computing the flow within a soap bubble on a staggered spherical grid~\cite{HuangSoapBubblesSIGGRAPH2020}.
	On evolving surfaces, Mercier et al. \cite{mercier2015surface} solved a PDE in a 3D narrow-band grid using the CPM to enhance the original simulation with additional turbulence.  Closest to our approach is the Semi-Lagrangian Closest Point Method \cite{auer2013semi}. In comparison to existing CPM-based methods, we add equations for conservation of mass. This allows us to correctly adapt scalar values when the surface area changes. We also adjust vector quantities to account for surface changes. Furthermore, we model a one-way coupling using mass and momentum transfer to create a plausible linkage between the evolution of the input surface and the behavior of the resulting 2D simulation.
	
 %and the rendering approach from Glassner \cite{glassner2000soap}. 

	%\section{Nomenclature}
	%\begin{center}
%		\begin{tabular}{cp{6cm}c}
%			Notation&&Unit\\
%			$M$& Material Surface &$m^2$\\
%			$\mathbf{n}$& Surface normal &\\
%			$dM$&Material Surface Element&$m^2$\\
%			$d\mathbf{M}$&Vector valued Material Surface %Element&$m^2$\\
%			$\mathbf{u}$&External velocity (Velocity of the %fluid/Surface/Isosurface)&$\frac{m}{s}$\\
%			$\mathbf{v}$&Velocity of the material on the %surface&$\frac{m}{s}$\\
%			$f$&Density/Concentration&?$\frac{1}{m^2}$\\
%			$\rho$&Mass density&$\frac{kg}{m^3}$\\
%		\end{tabular}
%	\end{center}
	%\cite{xu2003eulerian}
	
	\section{Modeling PDEs on Evolving Surfaces}\label{sec:model}

	In this section, we describe the fundamentals of our method and how we model the physics on the surface.
    We additionally describe the different physical phenomena that we model as secondary effects on the surface. 
	
	\subsection{Overview}
	Consider an evolving surface $M$. 
	Such a surface can result from a finite-difference or Smoothed Particle Hydrodynamics (SPH) simulation, a keyframe animation, or any time-dependent process.
	On $M$, we define another process (e.g., the simulation of a substance) with a set of PDEs and  couple it to the base animation $M$ results from.
	Therefore, the overall system is divided into three aspects:
	\begin{itemize}
		\item \textbf{Evolution of the surface $M$}:\\
		    Due to the surface evolution, the space on which we model the dynamic process is altered. 
		    In this step, we account for this fact, i.e., how quantities evolve alongside the surface.
	    \item \textbf{Coupling}:\\
            This aspect determines how the base animation affects the one on the surface.
		\item \textbf{Modeling the dynamics on the surface}:\\
		    This step accounts for the secondary dynamic process and how it is modeled on the surface, i.e., how a set of PDEs can be solved on the surface.
	\end{itemize}{}
	An example of such a process could be a drop of ink on a plastic sheet moving in space.
	The first aspect (evolution of the surface) describes how the ink is transported in space when the plastic sheet is moving in normal direction. 
	The second aspect (coupling) accounts for the movement of the plastic sheet in tangential direction, i.e., the acceleration of the ink due to friction forces, and the third aspect (the dynamic process on the surface) is concerned with the fluid behavior of the ink itself.

    \subsection{Evolution of the Surface}\label{susec:surf_evolve}
	To model such a dynamic process on an evolving surface $M\subset\mathbb{R}^3$, we need to define scalar as well as vector-valued quantities on it.
	To this end, we attach scalar quantities $a(\mathbf{p},t)\in\mathbb{R}$ and vectorial quantities $\mathbf{v}:=\mathbf{v}(\mathbf{p},t)\in T_{\mathbf{p}}M$ to every point $\mathbf{p}\in M$, where $T_{\mathbf{p}}M$ denotes the tangent space at point $\mathbf{p}$ defined by the surface normal $\mathbf{n}:=\mathbf{n}(\mathbf{p},t)$ and $t$ denotes a certain point in time.
	Typical quantities would be, e.g., mass density and velocity for fluid flow on the surface. 
    The velocity field $\mathbf{u}:=\mathbf{u}(\mathbf{p},t)$ defines how the surface evolves, i.e., how $M(t_0 + \Delta t)$ results from $M(t_0)$, where $\Delta t\in \mathbb{R}_{>0}$.
	For convenience, we will write $M$ instead of $M(t_0)$ and $M'$ for $M(t_0+\Delta t)$.
	In the following, the surface evolution characterized by $\mathbf{u}$ will be referred to as the outer process and the one on the surface will be called inner dynamics.
	
\begin{figure}[t]
	\includegraphics[width=0.9\linewidth]{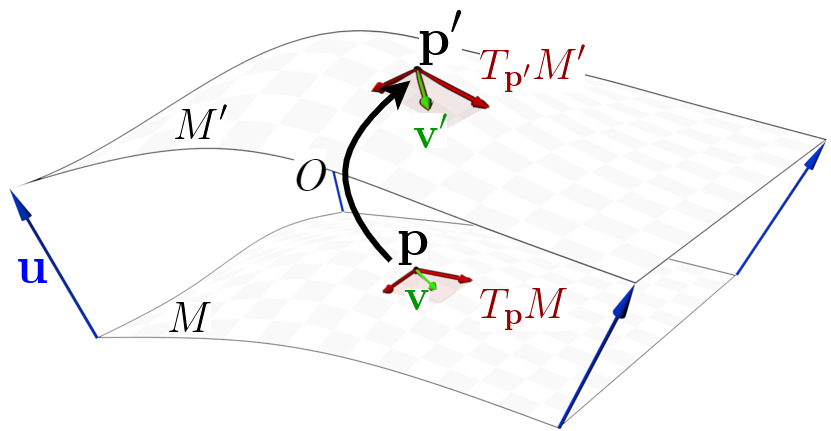}
	\caption{
	The map $O$ relates $\mathbf{p}\in M$ and $\mathbf{p}'\in M'$ and also maps tangent space $T_\mathbf{p}M$ into $T_{\mathbf{p}'}M'$.
	}
	\label{fig:illustration_of_O_v}
	\end{figure}

	We construct a map $O$ that describes the evolution of quantities due to the outer process.
	To this end, we define the space $\mathcal{M}$:
	\begin{align}
    	\mathcal{M} = \bigcup_{\mathbf{p}\in M} \{\mathbf{p}\}\times \mathbb{R} \times T_\mathbf{p}M\,.
	\end{align}
	The quantities needed for the inner dynamics can now be described as elements of $\mathcal{M}$.
	Without loss of generality, we consider only one scalar and one vectorial quantity attached to point $\mathbf{p}$. 
	It is possible to map an arbitrary number of quantities to $\mathbf{p}$ in the same way.
	The map $O$ relates elements from $\mathcal{M}$ to elements in space $\mathcal{M}' = \bigcup_{\mathbf{p'}\in M'} \{\mathbf{p'}\}\times \mathbb{R} \times T_{\mathbf{p}'}M'$:
	\begin{align}
	O:\mathcal{M} \rightarrow \mathcal{M}': \begin{pmatrix}\mathbf{p}\\a\\\mathbf{v}\end{pmatrix} \mapsto \begin{pmatrix}\mathbf{p'}\\a'\\\mathbf{v'}\end{pmatrix}.
	\end{align}
    An illustration of $O$ can be found in Fig.~\ref{fig:illustration_of_O_v}.
	
    We divide the influence of the outer process defined by $\mathbf{u}$ in normal $\mathbf{u}_n$ and tangential $\mathbf{u}_t$ components. 
	The tangential component is used to model friction between the outer and inner dynamics and discussed in Section~\ref{susec:coupling}.
	For now, we assume that the dynamics are coupled without friction and, therefore, $\mathbf{u}_t$ does not affect the inner dynamics, i.e., $m\in\mathcal{M}$ is only altered by $\mathbf{u}_\mathbf{n}$.
	
	% Position changes
	The rate of change $\frac{D_O}{Dt}$ that a point $\mathbf{p}$ experiences from to action of $O$ is then defined by $\tilde{\mathbf{u}}_\mathbf{n} := k\,\mathbf{u}_\mathbf{n}$, where $\frac{D_O}{Dt}$ denotes the material derivative and $k\in \mathbb{R}$ has to be chosen in such a way that $\mathbf{p}'\in M'$ holds.
	The motion of point $\mathbf{p}$ is directly governed by $\frac{D_O\mathbf{p}}{Dt}$, in particular its velocity is $\tilde{\mathbf{u}}_\mathbf{n}$ and, hence, $\frac{D_O\mathbf{p}}{Dt} = \tilde{\mathbf{u}}_\mathbf{n}$.
	
	%Scalar quantity changes 
	A scalar quantity $a$ is advected alongside the point $\mathbf{p}$. 
	We distinguish  between two types of scalar quantities: intensive and extensive ones~\cite{Redlich:1970:IntensiveExtensive}.
	An extensive property is a global property (e.g., mass or volume). 
	Such a property is only advected alongside $\mathbf{p}$ and not changed, i.e., $\frac{D_O a}{Dt} = 0$.
	In contrast, if $a$ describes an intensive physical property, such as density, we demand that 
	\begin{align}\label{eq:1}
	0 &= \frac{D_O}{Dt}\int_{M(t)}a(\mathbf{p},t)\,dM(\mathbf{p},t)
	\end{align}
	holds true, where $dM(\mathbf{p},t)$ are infinitesimal surface elements at position $\mathbf{p}$ on the surface $M$ at time $t$. 
	In the following, we will skip $t$ and $\mathbf{p}$ for convenience.
    Eq.~\ref{eq:1} ensures that the total amount of such a quantity does not change if no phenomena like mass transfer or chemical reactions are present.
	In this case, the rate of change of an intensive quantity $a$ is given by
	\begin{align}\label{eq:mass_cons_result}
	\frac{D_Oa}{Dt} &= - a(\nabla\cdot(\tilde{\mathbf{u}}_\mathbf{n})_{T_\mathbf{p}M})\,,
	\end{align}
	where $\mathbf{u}_{T_\mathbf{p}M} = (I - \mathbf{n} \mathbf{n}^T)\mathbf{u}$ is the velocity projected onto the tangent space $T_\mathbf{p}M$ at point $\mathbf{p}$.
	This means that, if the surface diverges, the concentration of $a$ decreases, and vice versa.
	A detailed derivation is provided in Appendix~\ref{sec:app1}.

\begin{figure}[t]
	    \centering
		\includegraphics[width=0.95\linewidth]{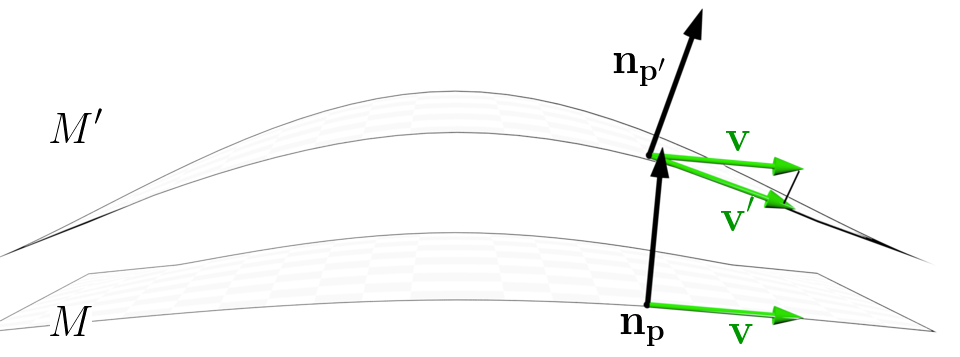}
		\caption{The map $O$ transforms velocity vector $\mathbf{v}$ to vector $\mathbf{v}'$. First, $\mathbf{v}$ is advected to the point $\mathbf{p}'$. Next, it is projected back into $T_{\mathbf{p}'}M'$ and scaled to conserve momentum.}
		\label{fig:two}
	\end{figure}

	%Vectorial quantity changes
	When advecting a vectorial quantity $\mathbf{v}$, we demand that $\mathbf{v}'\in T_{\mathbf{p}'}M'$ holds after applying $O$.
	This can be ensured by considering $\mathbf{v}$ to be a small linear material line element subjected to the velocity field $\tilde{\mathbf{u}}_\mathbf{n}$.
	Its rate of change is the difference of the velocities at the two ends of the element and can be described by $\frac{D_O\mathbf{v}}{Dt} = \nabla\tilde{\mathbf{u}}_\mathbf{n}\mathbf{v}$.
    In our model, the vector field $\mathbf{v}$ describes the velocity of the inner dynamics.
    As the directions of the velocities are constrained by the surface evolution, 
    the total momentum will be altered and, therefore, cannot be conserved.	
    We can only ensure that the sum of the absolute values does not change, i.e., we can enforce that
	\begin{align}\label{eq:abs_momentum_conservation}
		0 = \frac{D_O}{Dt}\int_M \norm{a\mathbf{v}}dM
	\end{align}
	will hold.
	Eq.~\ref{eq:abs_momentum_conservation} leads to 
	\begin{align}\label{eq:momentum_conservation_result}
		\frac{D_O\norm{\mathbf{v}}}{Dt} = 0\,.
	\end{align}
    Hence, we need to adjust $\frac{D_O\mathbf{v}}{Dt}$, so that the length of $\mathbf{v}$ is conserved.
    Using first-order Taylor expansion, the change of the length can be computed by $\mathbf{v}\cdot\nabla\tilde{\mathbf{u}}_\mathbf{n}\mathbf{v}\frac{\mathbf{v}}{\|\mathbf{v}\|^2}$.
	A detailed derivation can be found in Appendix~\ref{sec:app_length_preserving}.
	This term is then subtracted to adhere Eq.~\ref{eq:momentum_conservation_result}.
    Metaphorically, this means that $\mathbf{v}$ is advected and rotated back onto the surface.
    An illustration of the advection of a vectorial quantity is given in Fig.~\ref{fig:two}.

	If $a$ is an intensive property, the combined material derivative of the map $O$ then reads as 
		\begin{align}\label{eq:conservingO}
		\frac{D_O}{Dt}O =  
		\begin{pmatrix}
			\tilde{\mathbf{u}}_\mathbf{n}\\ 
			- a(\nabla\cdot(\tilde{\mathbf{u}}_\mathbf{n})_{T_\mathbf{p}M})\\
			\nabla \tilde{\mathbf{u}}_\mathbf{n}\mathbf{v} - \mathbf{v}\cdot\nabla\tilde{\mathbf{u}}_\mathbf{n}\mathbf{v}\frac{\mathbf{v}}{\|\mathbf{v}\|^2}
		\end{pmatrix}.
	\end{align}
   	
    Note that $\frac{D_O}{Dt}a$ denotes the rate of change of a quantity $a$ induced by the operator $O$ and $\frac{D_O}{Dt}O$ denotes the combined material derivative of $O$.
	
	 \subsection{Coupling}\label{susec:coupling}
	
%    So far, we ignored the tangential part $\mathbf{u}_t$ of $\mathbf{u}$.
    So far we assumed that the outer process influences the inner dynamics solely in normal direction.
	Next, we discuss how the tangential part $\mathbf{u}_t = \mathbf{u} - \mathbf{u}_n$ will influence the inner dynamics.
	This implies that friction-less coupling is modeled, i.e., matter slides on the surface and no adhesion is present.
	We model adhesive forces to reduce the relative velocities $\mathbf{v}^{\text{rel}} = \mathbf{v} - \mathbf{u}_t$ between the outer process and inner dynamics as
	\begin{align}\label{eq:couple0}
	\frac{D_c\mathbf{v}}{Dt} = -s_1\mathbf{v}^{\text{rel}}\,,
	\end{align}
	where $s_1$ is a user-defined coefficient and $\frac{D_c}{Dt}$ denotes the rate of change induced by coupling effects. 
    Similar to Section~\ref{susec:surf_evolve}, we could also consider $\mathbf{v}$ to be a small linear material line element exposed to the velocity $\mathbf{u}_t$ and its rate of change would then read as $\left(\nabla_{T_\mathbf{p}M}\mathbf{u}_t \right)\mathbf{v}$, where $\nabla_{T_\mathbf{p}M}$ is defined as $\nabla_{T_\mathbf{p}M} = (I-\mathbf{n}\mathbf{n}^T)\nabla$.
    We combine both views and model the total rate of change of $\mathbf{v}$ caused by coupling effects as
	\begin{align}\label{eq:couple1}
	\frac{D_c\mathbf{v}}{Dt} = -s_1\mathbf{v}^{\text{rel}} + s_2 \left(\nabla_{T_\mathbf{p}M}\mathbf{u}_t \right)\mathbf{v}\,.
	\end{align}
% 	In general, $s_2$ should be chosen to equal $s_1$. 
% 	However, it might be useful to have individual control in some cases.

	The outer process induces sinks and sources to the inner dynamics modeled on the surface.
	Therefore, we model transfer of intensive physical quantities (e.g., density) from the outer process to inner dynamics via sinks and sources on the surface.
	Inspired by Fick's laws of diffusion, we want that the concentration $a^{\text{out}}$ from the outer process and the concentration $a$ of the inner dynamics will align. 
	In other words, the concentration of a substance on the surface will be changed by
	\begin{align}\label{eq:couple2}
		\frac{D_c a}{Dt} = - s_3 a^\text{rel}\,, 
	\end{align}
	to reduce the concentration difference $a^\text{rel} = a - a^{\text{out}}$. 
	The speed of the reduction is determined by the factor $s_3$. %$s_3\in\left[0,1\right]$. 
	If only sources are modeled, the concentration difference is restricted to negative values, i.e., $a^\text{rel}\leq0$.
	Sinks are modeled by restricting $a^\text{rel}\geq0$.
	
	\subsection{Modeling the Dynamics on the Surface}\label{susec:2DSimulation}
 
	After discussing the surface evolution and coupling, we describe how dynamics on the surface can be modeled.
	To describe such a process we use a set of PDEs, e.g.,
  	\begin{align}
		\frac{D_s \mathbf{v}}{Dt} &= F_1\left(t,a,\mathbf{v},
		\partial_{1}a,
		\partial_{2}a, ...,
		\partial_{1}\mathbf{v},
		\partial_{2}\mathbf{v},...\right)\,\text{ and }\label{eq:sim_velocity_model}\\
		\frac{D_s a}{Dt} &= F_2\left(t,a,\mathbf{v},
		\partial_{1}a,
		\partial_{2}a, ...,
		\partial_{1}\mathbf{v},
		\partial_{2}\mathbf{v},...\right)\,,\label{eq:sim_density_model}
	\end{align}
	where the functions $F_1$ and $F_2$ characterize the phenomena modeled on the surface.  
	The term $\partial_{i} := 
	\left\{\frac{\partial^{|\boldsymbol{\alpha}|} }{\partial x_1^{\alpha_1}\partial x_2^{\alpha_2}\partial x_3^{\alpha_3}} :\,|{\alpha}|=i \right\}$ 
	denotes the set of all partial derivatives of the order $i$, with 
	${\boldsymbol{\alpha}}$ being a multi-index.
	For example, if fluid flow is modeled, $F_1$ describes the momentum conservation and $F_2$ the mass conservation of the Navier-Stokes equations.
	
	To solve Eqs.~\ref{eq:sim_velocity_model} and~\ref{eq:sim_density_model} on the surface, we only allow for tangential deviations.
	To this end, we solve them locally in the corresponding tangent spaces $T_\mathbf{p}M$.
	This implies that, instead of using the ordinary spatial derivations, we project them onto the surface, e.g., the operator $\nabla$ is replaced by $\nabla_{T_\mathbf{p}M}$.
	The governing equations for our model are then given by:
	\begin{align}
	\frac{D\,\mathbf{v}}{Dt} &= \frac{D_O\mathbf{v}}{Dt} + \frac{D_c\mathbf{v}}{Dt} + \frac{D_s\mathbf{v}}{Dt}\, \text{ and }\label{eq:gov_vel}\\
	\frac{D\,a}{Dt} &= \frac{D_Oa}{Dt} + \frac{D_ca}{Dt} + \frac{D_sa}{Dt}\,.\label{eq:gov_dens}
	\end{align}
	While the outer process is ignored in Equations~\ref{eq:sim_velocity_model} and~\ref{eq:sim_density_model}, it is included in Equations~\ref{eq:gov_vel} and~\ref{eq:gov_dens}, i.e., they govern the overall dynamics on the surface.
    
    Next, we describe different physical phenomena that we use in this paper to show the versatility of our model.
	
	\subsection{Examples}\label{sec:example_models}
	We model different physical phenomena on the surface and, therefore, employ different sets of equations.

	\begin{figure}[t]
		\centering
		\includegraphics[width=\linewidth]{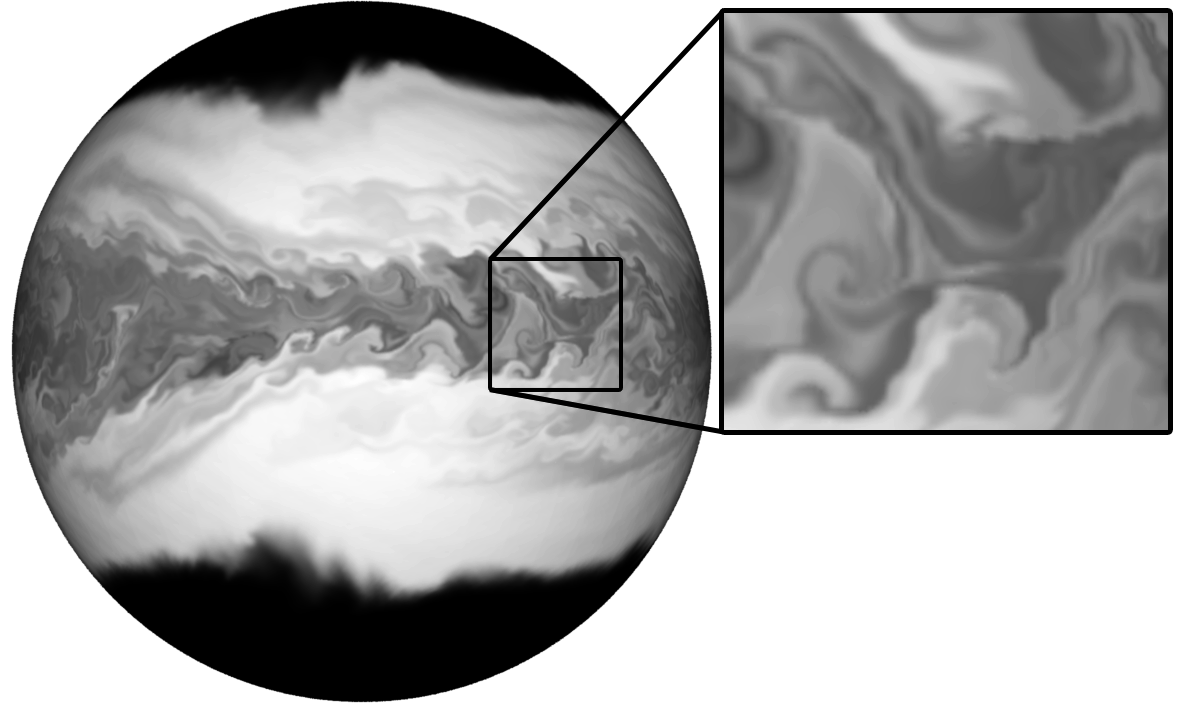}
		\caption{
		A rotating sphere where we couple the velocity of the internal simulation with the velocity of the sphere.
		Details are added with artificial vorticity and coupling noise. Simulation resolution is 207 $\times$ 207 $\times$ 207.
		}
		\label{fig:rotating_sphere}
	\end{figure}
\begin{figure}
	    \centering
		\includegraphics[width=\linewidth]{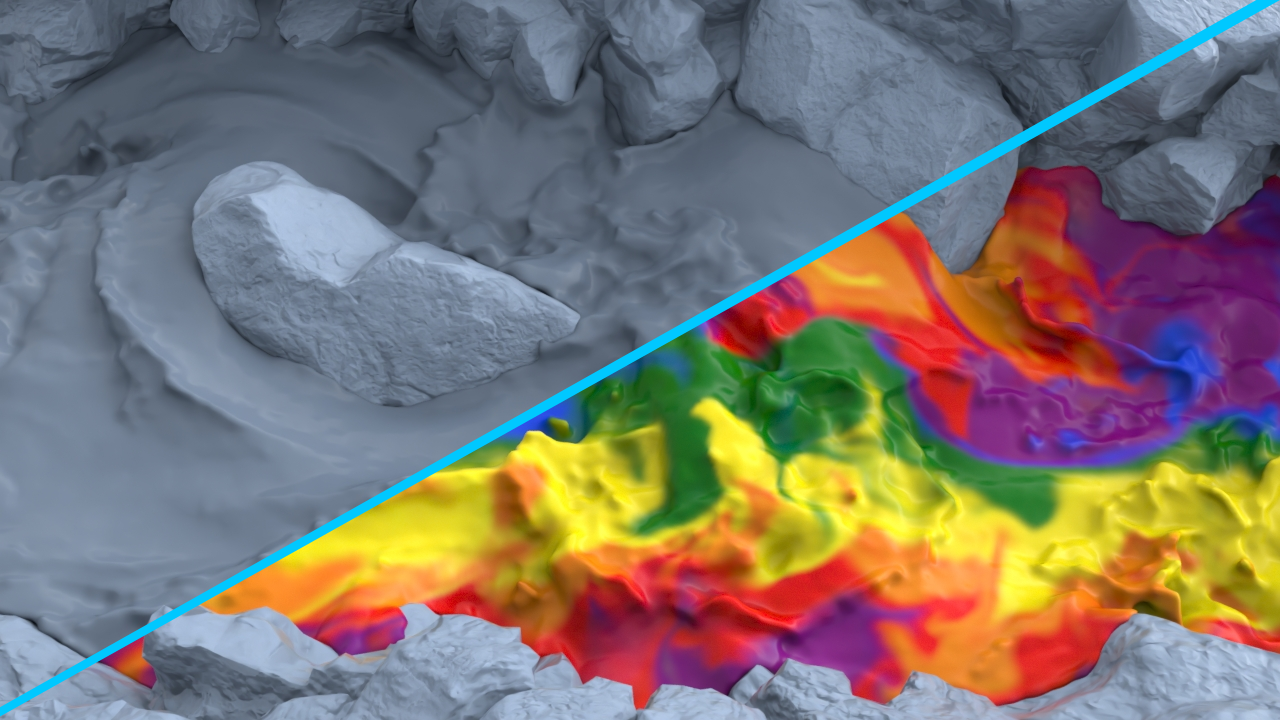}
		\caption{Fluid flow simulation on top of a FLIP fluid riverbed simulation. Performing a surface simulation of a color-band demonstrates that we can add fine-scaled details. The split screen shows the input fluid simulation on the left and the added surface flow on the right.}
		\label{fig:river}
	\end{figure}
\begin{figure}[t]
    \centering
		\subfloat[Reaction-diffusion]{\includegraphics[width=.8\linewidth]{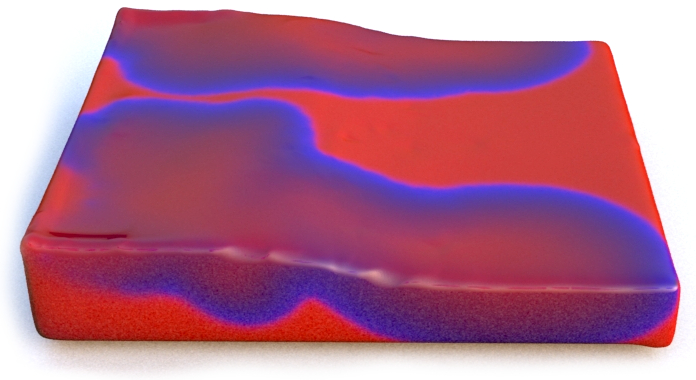}\label{fig:dam}}
		\\
		\subfloat[Fluid flow]{\includegraphics[width=.8\linewidth]{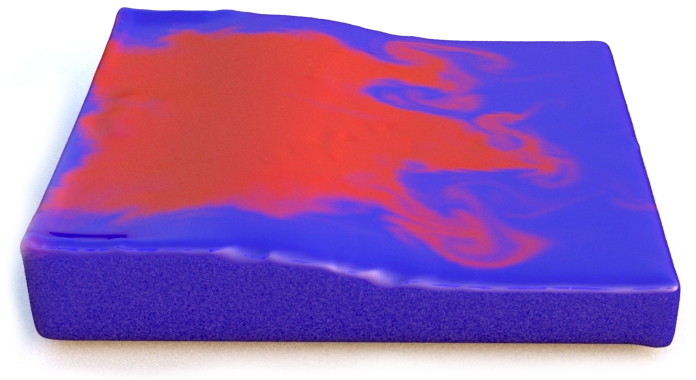}\label{fig:dam_fluid}}
		\caption{
		With our approach, we can model different behaviors on the same input data.
		The upper image shows a reaction-diffusion simulation on top of a dam break SPH simulation. The lower image shows a fluid flow on top of the same input simulation. Simulation resolution is 199 $\times$ 50 $\times$ 169.
		}
	    \label{fig:dam_break}
	\end{figure}

	\paragraph*{Fluid Flow.}
	One application of our model is the simulation of fluid flow on an evolving surface. 
	Figs.~\ref{fig:rotating_sphere},~\ref{fig:river}, and~\ref{fig:dam_fluid} show examples of such a flow. 
	The surface rotates and, due to friction forces, the fluid on the surface starts to move.
    To model viscous fluid flow on the surface we use the Navier-Stokes momentum equation:  
	\begin{align}\label{eq:fluid_momentum_equation}
		\frac{D_s\mathbf{v}}{Dt} = - \frac{1}{\rho}\nabla_{T_\mathbf{p}M} P + \nu {\nabla^{2}_{T_\mathbf{p}M}}\mathbf{v} + \frac{1}{\rho}\mathbf{F}_b\,,
	\end{align}
	where $\mathbf{v}$ is the fluid's velocity, $\nu$ the viscosity constant, $P$ is the pressure, and $\mathbf{F}_b$ are body forces (e.g., gravity).
	
	The evolution of density $\rho$ is modeled by using the general continuity equation:
	%In its differential form, it states that the change $\frac{D\,\rho}{Dt}$ of density $\rho$ of the fluid can be defined as
	\begin{align}\label{eq:continuity_equation}
		\frac{D_s\rho}{Dt}  = \sigma - \rho(\nabla_{T_\mathbf{p}M}\cdot\mathbf{v})\,,
	\end{align}
	where $\sigma$ is the rate of generation of the substance per unit volume, i.e., it is used to model sinks ($\sigma<0$) or sources ($\sigma>0$).
	We model incompressible fluid flow on the surface and only allow for density changes due to surface divergence, i.e., we set $F_2(\rho,\mathbf{v}) = 0$.
	As a result, Eq.~\ref{eq:continuity_equation} simplifies to a volume conservation law:
	\begin{align}\label{eq:continuity_equation_incomp}
	  \nabla_{T_\mathbf{p}M}\cdot\mathbf{v} = \frac{\sigma}{\rho}\,.
	\end{align}	
	If no sinks or sources are present (i.e., $s_3=0$), Eq.~\ref{eq:continuity_equation} simplifies to $\nabla_{T_\mathbf{p}M}\cdot\mathbf{v} = 0$.
	Note that if the outer process is divergence-free, i.e., $\nabla\cdot\mathbf{u} =0$, we obtain an incompressible flow because $\frac{D\,\rho}{Dt}=0$.

	    \begin{figure}[t]
		\includegraphics[width=\linewidth]{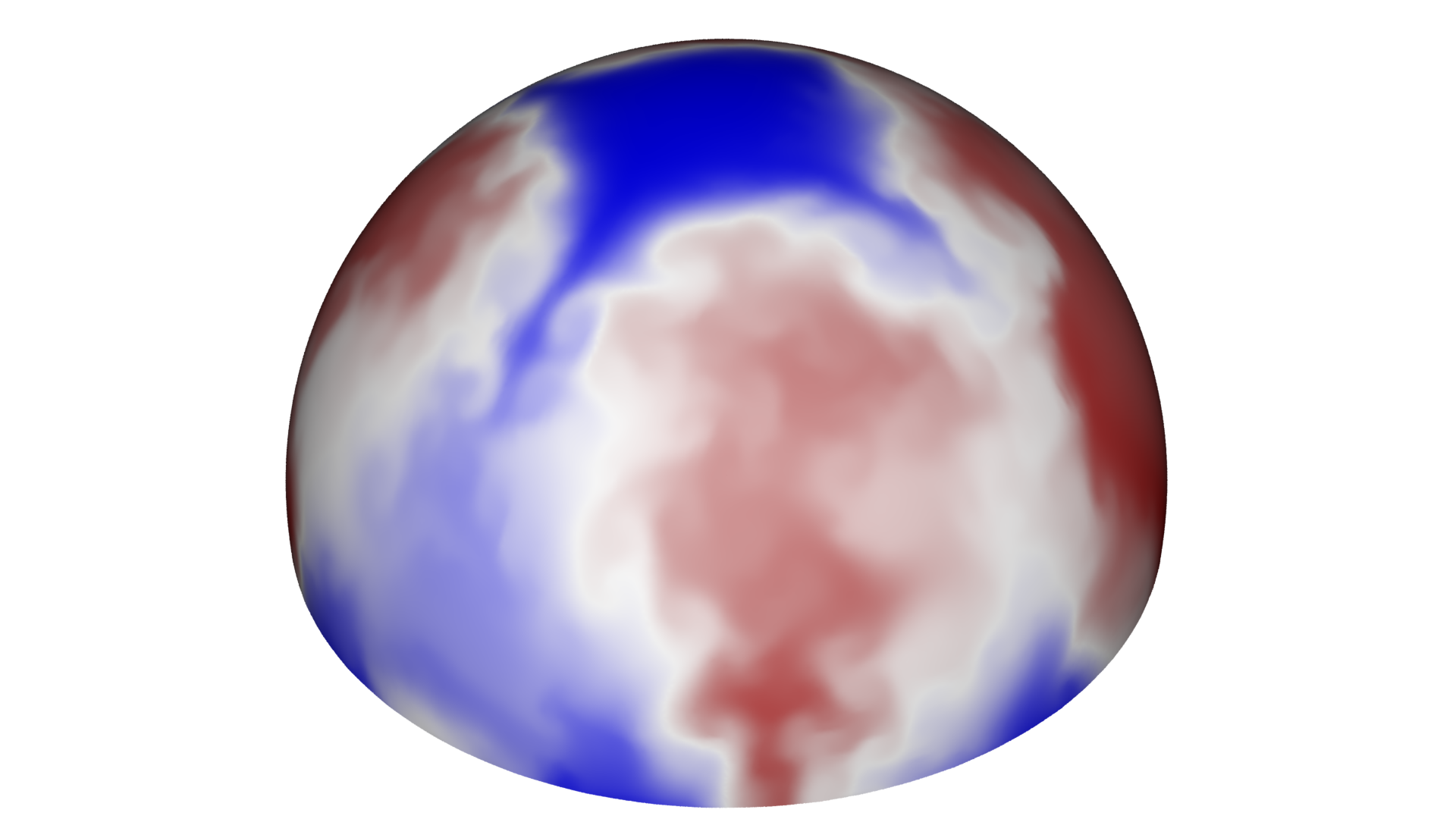}
		\caption{
		Thermal convection on a hemisphere.
		The temperature is color-coded, where the temperature rises from blue over white to red.
		Some areas are heated and others are cooled. 
		Due to these temperature differences, buoyancy-driven flow arises. Simulation resolution is 207 $\times$ 107 $\times$ 207.
		}
		\label{fig:heat}
	\end{figure}
	\paragraph*{Buoyancy-induced Flow.} 
    Fig.~\ref{fig:heat} shows a static hemisphere. 
    Some areas on the surface are heated, others cooled by outer temperature constraints. 
    We model temperature transfer according to Eq.~\ref{eq:couple2} and, therefore, the fluid on the surface changes its temperature.
    Natural convection arises due to temperature differences in the fluid.
    
    Such buoyancy-driven flow can be modeled using the so-called Boussinesq approximation~\cite{boussinesq1897theorie}.
    The Boussinesq approximation assumes incompressible flow and that variations of density only occur due to temperature differences: $\rho = \rho_0 - \beta\rho_0(\mathcal{T}-\mathcal{T}_0)$, where $\beta$ is the coefficient of thermal expansion, $\mathcal{T}_0$ the reference temperature, and $\rho_0$ the reference density. We assume an ideal gas and, therefore, $\beta = \frac{1}{\mathcal{T}_0}$. If we assume gravity as the only body-force present, Eq.~\ref{eq:fluid_momentum_equation} becomes 
    \begin{align}\label{eq:bouyancy_momentum_equation}
		\frac{D_s\mathbf{v}}{Dt} = - \frac{1}{\rho}\nabla_{T_\mathbf{p}M} P + \nu {\nabla^{2}_{T_\mathbf{p}M}}\mathbf{v} + \left(1-\frac{\mathcal{T}}{\mathcal{T}_0}\right)\mathbf{g}\,,
	\end{align}
    with $\mathbf{g}$ being the gravitational constant.
    To model changes in the temperature field we use the convection-diffusion equation:    	\begin{align}\label{eq:thermal_convection_diffusion}
		\frac{D_s \mathcal{T}}{Dt} = \mu_d \nabla^2_{T_\mathbf{p}M} \mathcal{T}\,,
	\end{align}
	where we assume a constant diffusion coefficient $\mu_d$.
%	In Fig.~\ref{fig:heat} an example simulation of buoyancy induced flow on a half sphere is shown. 
%	To generate flow, we added heat sinks and sources at fixed positions on the sphere using Eq.~\ref{eq:couple2}.

	\paragraph*{Reaction-diffusion.}
	Reaction-diffusion is commonly used to model chemical reactions of one or more substances (e.g., a substance is transformed into another due to chemical reactions) and the diffusion of the substance(s) in space.
	Fig.~\ref{fig:dam} shows such a process. Two chemical substances diffuse and react with each other. 
	In this example, we model a two-component reaction-diffusion process given by
	\begin{align}\label{eq:react1}
	    \frac{D_sf_1}{Dt} &= R_1(f_1,f_2) + \mu_{d_1}\nabla^2_{T_\mathbf{p}M} f_1 \, \text{ and}\\
	    \frac{D_sf_2}{Dt} &= R_2(f_1,f_2) + \mu_{d_2}\nabla^2_{T_\mathbf{p}M} f_2\,.
	\end{align}
	The functions $R_1$ and $R_2$ are the reaction terms and characterize the system.
	
%	Turing~\cite{Touring52ReactionDiffusion} proposes the following reaction terms to model how different morphogens react:
%	A possible reaction term is given by Touring~\cite{Touring52ReactionDiffusion} as 
% 	\begin{align}\label{eq:react2}
% 	    R_1(f_1,f_2) &= \frac{1}{16}(16 - f_1f_2)\, , \\
% 	    R_2(f_1,f_2) &= \frac{1}{16}(f_1f_2 - f_2 - \beta)\, , 
% 	\end{align}
% 	where $\beta$ is a per-cell random number.
% 	Turk~\cite{turk1991generating} proposes choosing $\beta \in \left[11.95,12.05\right]$.
 
    We use the Gray Scott Model~\cite{GrayScott} to define the reaction terms, i.e., to model how different morphogens react:
    \begin{align}
	    R_1(f_1,f_2) &= - f_1f_2^2 + \beta(1-f_1)\, , \\
	    R_2(f_1,f_2) &= f_1f_2^2 - (\beta+\gamma)f_2\, ,\label{eq:react2} 
	\end{align}
	To generate the example illustrated in Fig.~\ref{fig:dam}, we set the parameters to $\beta=0.03$, $\gamma=0.06$, $\mu_{d_1}=0.1$, and $\mu_{d_2}=0.1$.

	\section{Connecting to the Base Animation}
    In this section, we describe how we apply the model from Section~\ref{sec:model} to create a simulation system. Our method can be divided into seven steps, as illustrated in Fig.~\ref{fig:schematic}. We start with a coarse input (Fig.~\ref{fig:schematic}, Step 1), e.g., a fluid simulation. Then, we create a signed distance field from this simulation (Fig.~\ref{fig:schematic}, Step 2) and write the simulation properties such as velocity, density, or color into 3D fields that are laid out in a narrow-band grid, which is our generic input format (Fig.~\ref{fig:schematic}, Step 3). We bring quantities from the outer process into our surface domain in a coupling step and compensate for the surface evolution as described in Section~\ref{subsec:surface_evolution} (Fig.~\ref{fig:schematic}, Step 4 and 5). Then, we simulate 2D details to enhance the coarse input by solving a PDE in the 2D surface space  (Fig.~\ref{fig:schematic}, Step 6) that results in a velocity field that advects all grids in the last step (Fig.~\ref{fig:schematic}, Step 7). 
    
    To solve a PDE on a surface we use the Closest Point Method (CPM) \cite{ruuth2008simple} calculated in a 3D narrow-band.
    %The details of their method are explained in their paper \cite{ruuth2008simple}. 
    The main idea of the CPM is that if the values of a field are constant along the normal of the surface, many equations calculated in this 3D field are the same as if they were calculated in the tangent spaces of the surface. 
    %Our examples are based on scalar fields, velocity fields, and gradients of them and, therefore, all fall into the category of equations that we can solve using the CPM. 
    Section~\ref{subsec:narrow} describes the process of converting 3D fields into fields that have constant values along the surface normal.
    
    We implemented our method where each part is interchangeable using a system of modules that change data flowing through the system. In the following, we will describe each step in detail.  
    
    \subsection{Data Input}
    The first step implements data input and is able to process a variety of base animation types. In our examples, we consider keyframe animation (Fig.~\ref{fig:rotating_sphere}), SPH particle simulation (Fig.~\ref{fig:coffee} and \ref{fig:dam}), and fluid implicit particle (FLIP) simulation (Fig.~\ref{fig:river}). The only requirement for an input is that we can convert the surface geometry data into a signed distance field and that values for the surface velocity can be generated on the surface.   
   
   \subsection{Convert to Signed Distance Field}
    For converting polygon meshes to signed distance fields, several methods are available, e.g., \cite{Sud2006}, \cite{xu2014signed}, or \cite{Koschier2017}. A special case arises when we convert particle systems to distance fields. Inside the fluid, there are particles everywhere. Here, the distance field obtained by these methods is not usable since we need the distance to the surface and not to the nearest particle. For particle-based simulations, we can either transform the simulation to a polygon mesh first or go directly from particles to signed distance fields by defining an implicit surface from the particles \cite{ZhuBridson2005} and then take the distance to this implicit surface.

    \subsection{Create Narrow-Band Grid}\label{subsec:narrow}
\begin{figure}[t]
        \centering
		\includegraphics[width=0.95\linewidth]{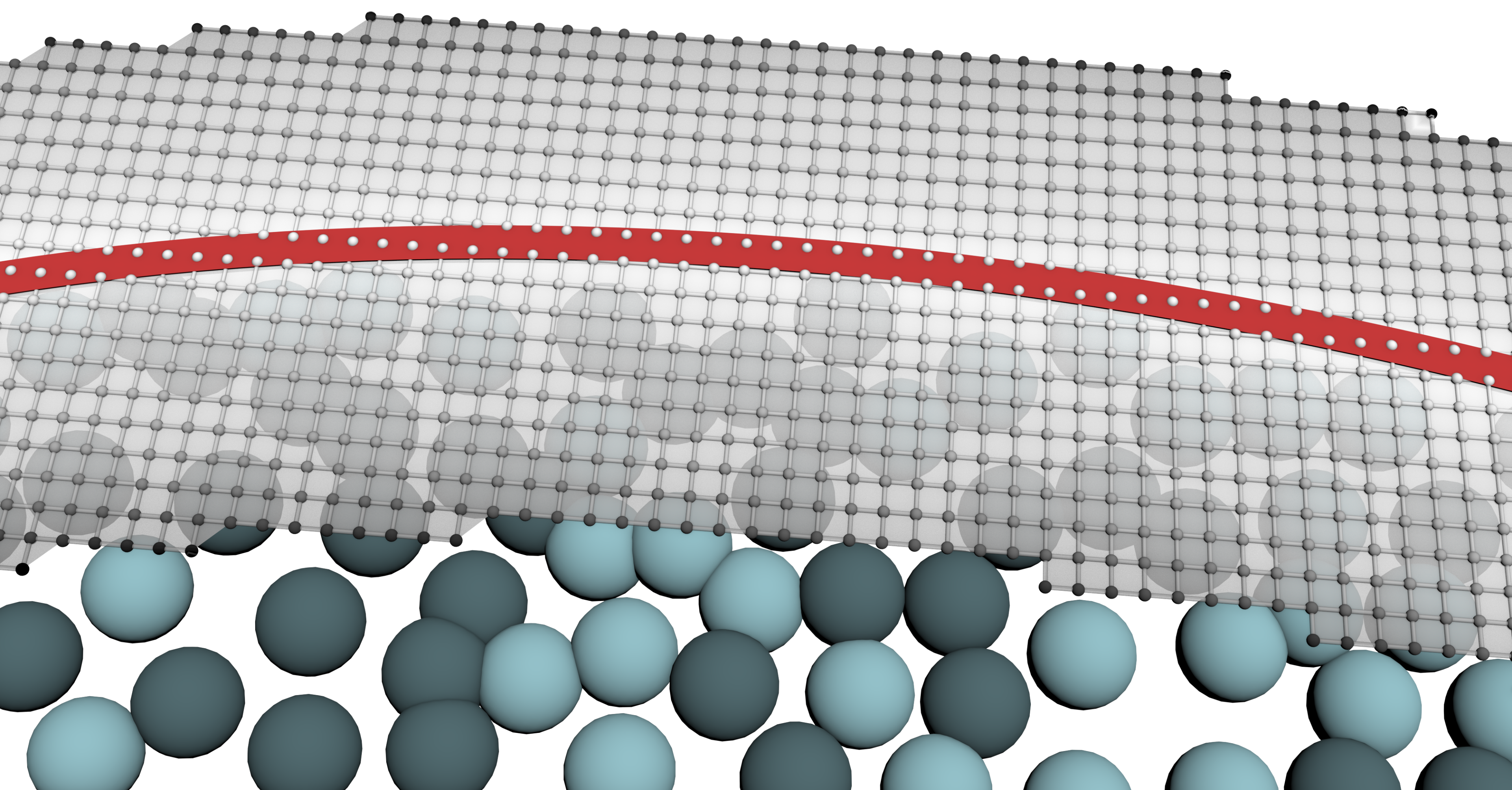}
		\caption{The distance field is stored in a narrow-band around the surface. In the same way, velocity and scalar fields are stored in a sparse data structure.
		\vspace*{-5pt}}
		\label{fig:distance}
	\end{figure}

    We create a narrow-band around the surfaces to capture the velocity, density, and other properties of the flow from the nearest surface point. We only fill cells near the surface based on the distance field. Cells that a farther away than a certain threshold are left empty, as illustrated in Fig.~\ref{fig:distance}. We keep track of two velocity fields, the internal velocity field (denoted as $\mathbf{v}$ in Section~\ref{sec:model}) and the external velocity field (denoted as $\mathbf{u}$). We write the velocity of the incoming surface into the external velocity grid. Depending on the use case, we either initialize the internal velocity grid with a snapshot of the external velocity (for example, Fig.~\ref{fig:coffee}), or we create a custom initialization routine to define the initial internal velocity, for example, by initializing with vanishing velocity (Fig.~\ref{fig:rotating_sphere}). We also keep track of the scalar properties that we use in our simulation, e.g., the concentration of substances for the reaction-diffusion example.
 
    To use the CPM our inputs must be narrow-band grids where the values are constant along the normal direction of the surface. Using an integer look-up grid that stores the cell coordinates of the closest point of the surface in each grid cell, we can fill in a grid with values from the closest surface point. We refer to this step as the CPM extension. Here, we extend the values near the surface along the normal direction. %, as can be seen in Fig.~\ref{fig:cptextension}.
    The CPM extension is our final step in the narrow-band creation phase. 
    
    %\begin{figure}[t]
	%	\begin{overpic}[width=\linewidth]{images/cptextension.png}
	%	    \put (36.5,30) {CPM Extension}
	%	\end{overpic}
	%	\caption{The CPM extension step extends the values closest to the surface along the normal direction. The red line indicates the surface. The blue color is a scalar value in the volume. On the left is the input field. On the right is the resulting CPM extension.}
	%	\label{fig:cptextension}
	%\end{figure}

    \subsection{Surface Evolution} \label{subsec:surface_evolution}
    Before we use the velocities or scalar values like color or density in a PDE, we first execute the correction steps for density and velocity, as described in Section~\ref{sec:model}. 
    
    For mass conservation, we have to solve Eq.~\ref{eq:mass_cons_result}. As a building block for this calculation, we need a module that implements the operator $\nabla\cdot(\tilde{\mathbf{u}}_\mathbf{n})_{T_\mathbf{p}M}$. This operator will calculate a scalar value for the divergence of the velocity field but with respect to the tangent space of a surface defined by the gradient of the input distance field. The result is then applied to the scalar fields that need correction. The interesting part of this divergence calculation is that we use the original 3D velocities for the divergence operator. We need to take the velocities of adjacent cells into account but projected to the surface normal of the current grid value, not the normals of the respective neighbor cells. An example to better understand the difference is the one of a growing sphere. Although the 2D velocities in surface space are zero at each point, the divergence is not. To get correct 2D divergence where a growing sphere causes sinks, and a shrinking sphere creates sources in the mass conservation equation, we have to project the velocities of neighbor cells using the normal of the current cell.
    
    For the conservation of momentum, the \enquote{Project Vector} module alters the velocities based on the surface tangent. Given a source vector grid and a distance field gradient, this module will project the input vectors onto the surface by subtracting the normal vector component from the gradient field. There is an option to maintain the length from the input vector, which resembles rotating the vector down onto the surface, as shown in Fig.~\ref{fig:two}. 
    To apply Eq.~\ref{eq:couple1}, we add a module to calculate the jacobian of a velocity field and use it to change the velocities of a second velocity field. Both modules are applied to the velocity field for the internal velocity. 
    
    After the values are adapted based on the surface evolution, we couple them to the values from the underlying simulation.
    
    \begin{figure}[t]
		\includegraphics[width=\linewidth]{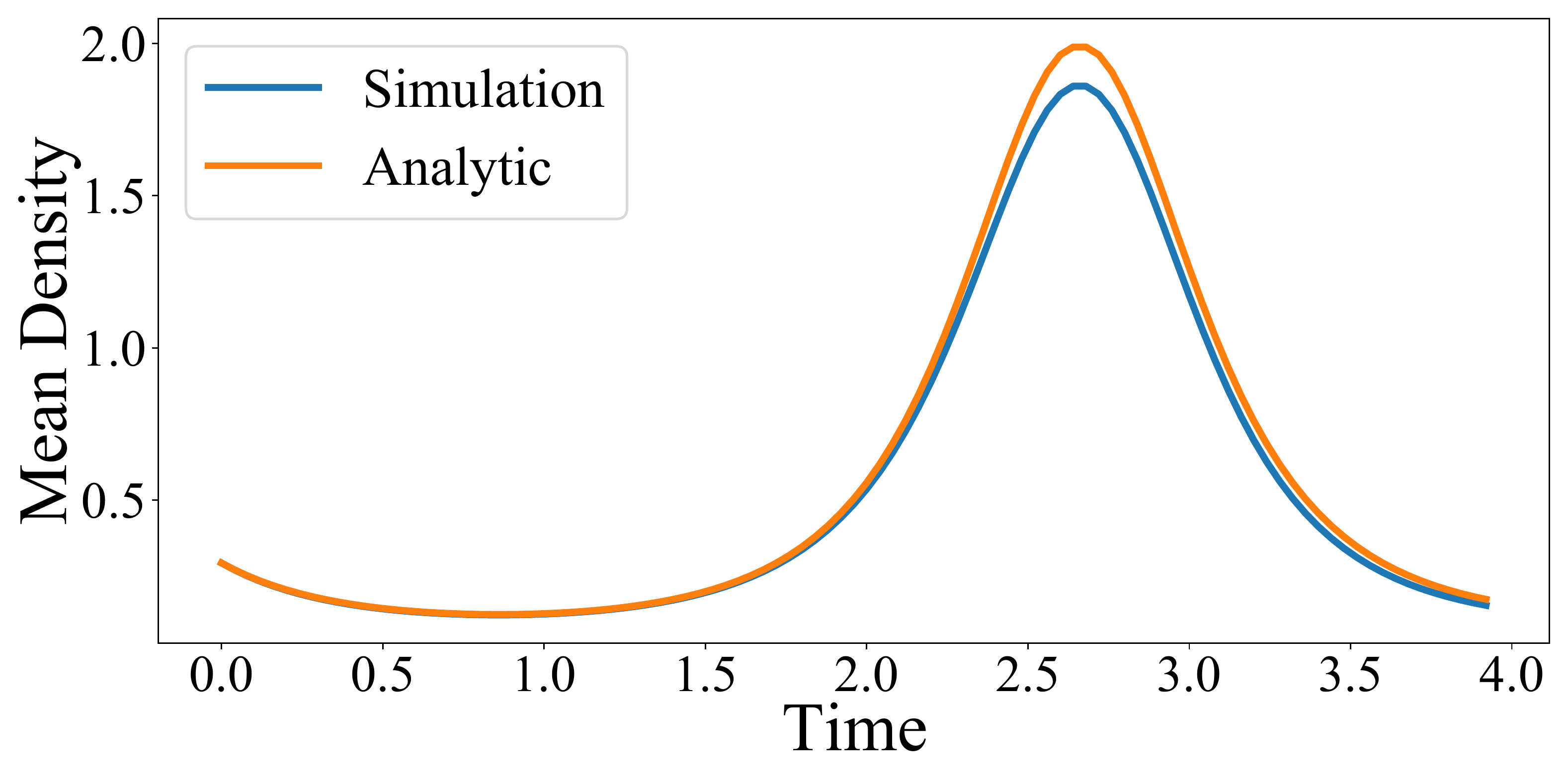}
		\caption{Simulated density on a growing and shrinking sphere compared to the analytic solution.
		With our simulation, we can reproduce the analytical result. 
		The numerical diffusion is negligible.}
		\label{fig:sphereplot}
	\end{figure}
    \subsection{Coupling of Dynamics}
%  	 \begin{figure}
% 		\includegraphics[width=\linewidth]{images/comparison.png}
% 		\caption{Comparison for different parameters for our coupling operator.}
% 		\label{fig:comparison}
% 	\end{figure}
    For coupling the velocities, we send the velocity grids to a module that applies Eq.~\ref{eq:couple1} to the inner velocity. The parameters for $s1$ and $s2$ are exposed to the user. We implemented functionality to add noise, viscosity, and vorticity to the inner velocity. 
    These parameters allow us to improve the coupling with additional details, as can be seen in Fig.~\ref{fig:rotating_sphere}.
    
    By coupling scalar values from the initial 3D simulation to the 2D space using Eq.~\ref{eq:couple2}, we can model mass transfer. The user can choose the parameter $s3$ to drive how much the outer process influences the values in the 2D simulation. A value of 0 would only initialize the 2D simulation, and from then on, the 2D simulation would be independent.   
    
    \subsection{Solve PDE and Advect}
    As mentioned, we are operating on 3D grids where values do not change along the normal direction. The CPM allows us to solve PDEs in the surface space using these grids. We implemented different sets of PDEs that can all be solved using our new modules. The modules work in 3D, but we specifically designed them for the fields that result from the CPM extension. 
    Due to the CPM, the gradient naturally operates in surface direction, but the divergence, curl, and Jacobian have to be changed: When applying them, we use the projected version as described in Section~\ref{sec:model}, i.e., the input vector is projected onto the surface.

    To simulate fluid flow on the surface, we solve Eq.~\ref{eq:fluid_momentum_equation}. 
    Here, we use the projected divergence operator $\nabla\cdot(\tilde{\mathbf{u}}_\mathbf{n})_{T_\mathbf{p}M}$.
    The \enquote{Divergence-Free} module will remove divergence with respect to the tangent space of a surface and take the divergence of the surface evolution into account.
    For the buoyancy-induced flow, we scaled the gravity depending on temperature as described in Eq.~\ref{eq:bouyancy_momentum_equation} and added a diffusion module for temperature diffusion. 
    To simulate reaction-diffusion for the example shown in Fig.~\ref{fig:dam} we wrote a module to implement Equations~\ref{eq:react1}--\ref{eq:react2}. 
    %The last step is the advection step where we calculate the advection of scalar and vector grids given a velocity grid.
	As a last step, we advect all fields using Eq.~\ref{eq:gov_vel}, i.e., with the resulting velocities.

	\subsection{Implementation}
    
	Our modular approach can be easily implemented with existing frameworks like the SideFX\texttrademark Houdini software package, which is widely used for VFX creation and offers an integration of the OpenVDB libraries \cite{Museth:2013:OOD:2504435.2504454} as a sparse volume representation for the calculations. We separated our algorithm into independent modules and implemented them as a separate operators, to be used in Houdini's simulation node graph framework. The OpenVDB framework offers a data structure, the OpenVDB grid, to create narrow-bands. All subsequent calculations employ this structure and only spend computation time where needed.  
    Our \enquote{CPM Extension} module is the central block to build CPM solutions in Houdini. Here, we implemented a nearest-neighbor interpolation as suggested by Kim et al. \cite{kim2013closest}, but also box and quadratic interpolation.  
    To generate the SPH base simulations, we used divergence-free SPH~\cite{Bender:2015:DFSPH} with consistent Shepard interpolation~\cite{Reinhardt:2019:Consistent}. For more details about SPH, we refer the reader to the SPH tutorial by Koschier et al.~\cite{Koschier:2019:SPHTutorial}. When simulating fluid flow, we used vorticity confinement ~\cite{fedkiw:2001:vorticity}.

\section{Results}
\subsection{Versatility and Simulation Quality}

\begin{figure*}[t]
	\centering
	\null\hfill
	\subfloat[Source SPH particles]{\includegraphics[width=0.24\textwidth]{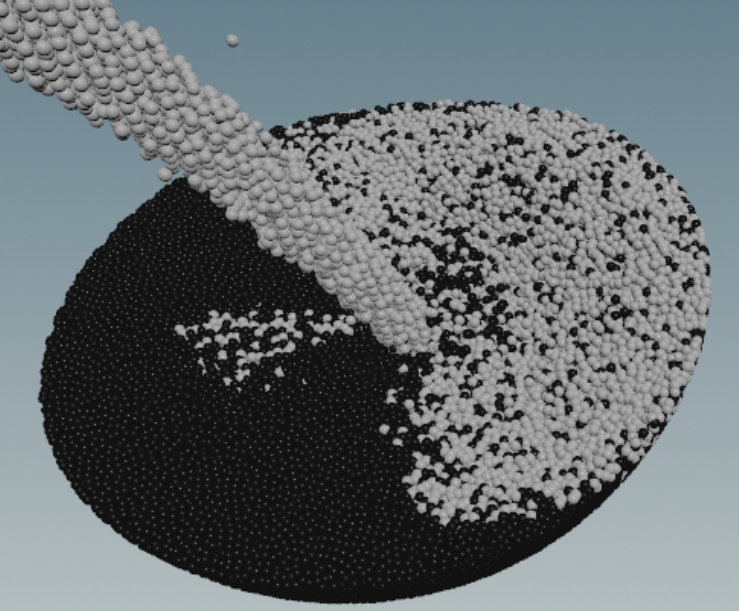}\label{fig:particles}}
	\hfill
	\subfloat[Original scalar field derived from particle attributes]{\includegraphics[width=0.24\textwidth]{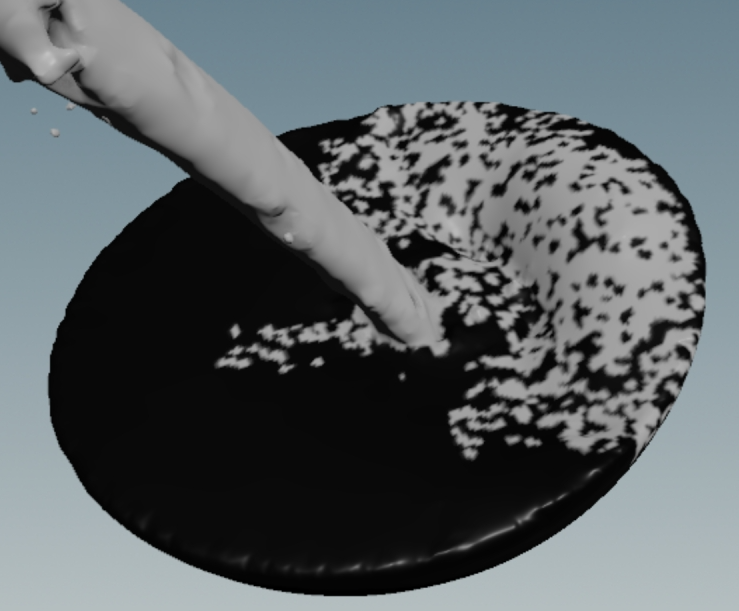}\label{fig:raw}}
	\hfill
	\subfloat[2D simulation with grid resolution of 0.05]{\includegraphics[width=0.24\textwidth]{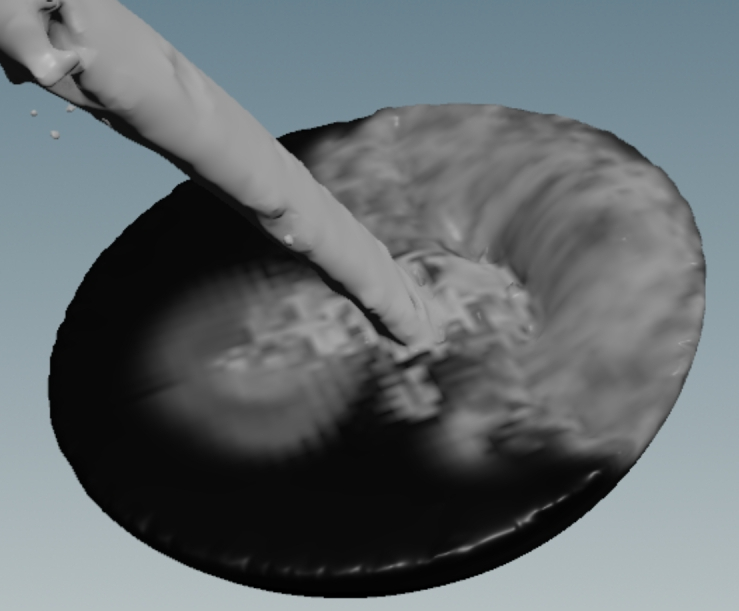}\label{fig:coffee_result}}
	\hfill
	\subfloat[2D simulation with grid resolution of 0.02]{\includegraphics[width=0.24\textwidth]{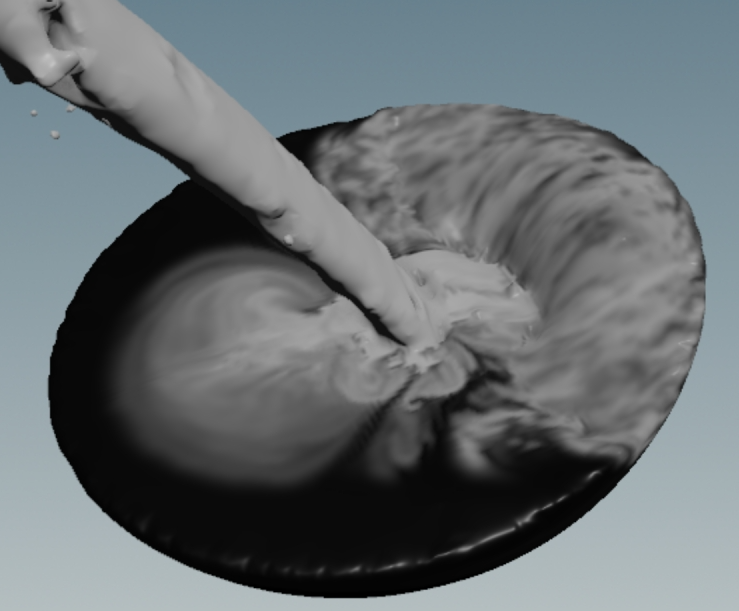}\label{fig:coffee_result3}}
	\hfill\null
	\caption{Pouring polluted water into a bowl. The fine-grained simulation enhances the underlying SPH simulation. Our mass transfer even captures oil particles that emerge from under the surface. Different resolutions can be generated independent from the input resolution. This allows iterative refinement of parameters as needed in typical VFX production workflows. }
	\label{fig:coffee}
\end{figure*}

To show the versatility of our method, we modeled different physical phenomena as described in Section~\ref{sec:example_models}, and tested them on a variety of scenarios.
We chose base animations with different characteristics, from static (Fig.~\ref{fig:heat}) to hand-animated meshes (Fig.~\ref{fig:rotating_sphere}), from coarse-scale SPH-based fluid simulations, including sudden changes in topology (Fig.~\ref{fig:dam_break}), to high-resolution multi-phase SPH simulations (Fig.~\ref{fig:coffee}).
In all of the above situations, we were able to add a fine-scale secondary simulation on top of these surfaces, revealing fine-scale details as promised.

With our approach, we can add effects onto the surface of the simulation.
Simulating a second fluid flow on the surface enables one to add another phenomena on top of an existing simulation. As shown in the oil film example (Fig.~\ref{fig:coffee}), the effect of oil spreading on the surface is achieved by simulating a second, fine-scale fluid flow on the surface, which is just added to the base-simulation.
As mentioned, the level of detail of the secondary simulation can be chosen independently and increasing  the simulation resolution of the underlying SPH simulation would not have the same effect. Instead, we would just have a scalar field with higher resolution, like the one shown in Fig.~\ref{fig:raw}, but would not have simulated the laws of the secondary flow.
Also, due to the modeled mass transfer, the oil particles that emerge from below the surface can contribute to our 2D simulation.
This coupling introduces significantly more details than just a plain simulation on the surface itself.
%This example also shows that we can create details in our narrow-band independent from the resolution of the input. 
The amount of detail added is significant even for low-resolution base animations, and it can be further improved (Fig.~\ref{fig:coffee_result} to~\ref{fig:coffee_result3}) by increasing the resolution. 
\begin{figure}[t]
	\includegraphics[width=\linewidth]{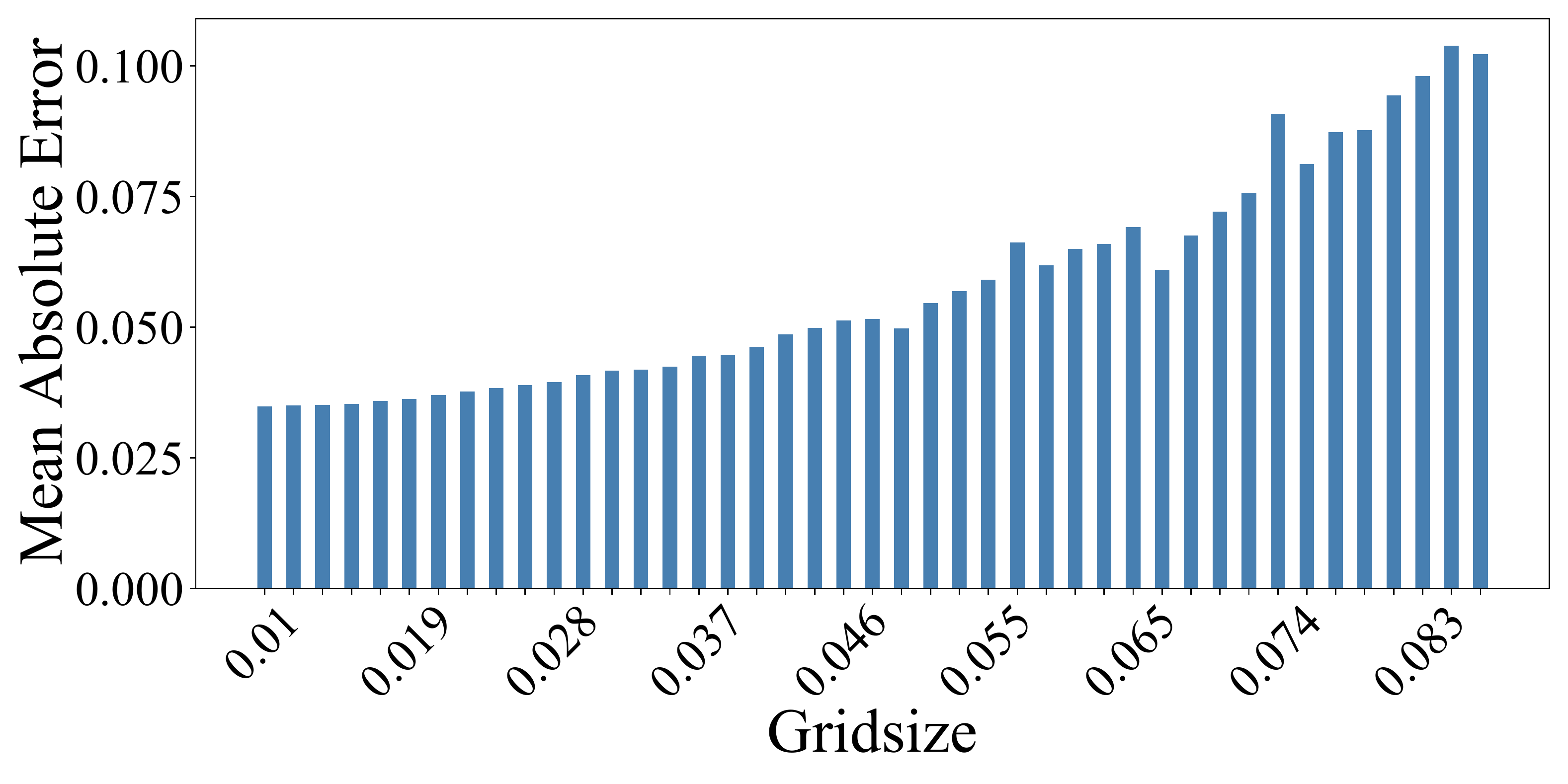}
	\caption{The mean error for different grid sizes. 
	The smaller the grid cells, the closer the values are to the analytic solution.
	}
	\label{fig:errorplot}
\end{figure} 
Fig.~\ref{fig:dam} presents a reaction-diffusion of two chemicals simulated on top of a coarse dam break simulation. This example illustrates that we can simulate not only a second flow equation on the surface but any kind of desired physical phenomena that can be described by a set of PDEs.

In the buoyancy-induced flow example (Fig.~\ref{fig:heat}), we simulate thermal convection and demonstrate emergence of isolated vortices on a static surface, replicating the physical experimental setup by Seychelles et al.~\cite{seychelles2008thermal}.
The setup consists of a static hemisphere on a heating plate giving rise to thermal convection. 
We were able to create fine detailed vortex structures using the Boussinesq approximation.

A hand-animated sphere with a periodically oscillating radius was used to test our approach on mass conservation. The mean density per surfactant was calculated and plotted as a function over time as a graph (Fig.~\ref{fig:sphereplot}). In this example, it is possible to analytically calculate the density change that is needed to ensure mass conservation and compare it with our simulated result. 
As shown in Fig.~\ref{fig:sphereplot}, we can reproduce the analytical solution over time.
The small loss in density can be attributed to numerical diffusion and is negligible in this case. 
To test the accuracy depending on the grid resolution, we simulated this scenario with different grid resolutions.
As can be seen in Fig.~\ref{fig:errorplot}, our mass conservation approach works as expected and the simulation converges to the analytical solution using finer grid resolutions. Further investigations on the convergence order is left for future work. 

%\begin{figure*}[t]
%	\centering
%	\null\hfill
%	{\begin{overpic}[width=0.33\textwidth]{images/velocity_conservation1.jpg}
%	 \put (22,3) {\color{black} a}
%	 \put (50,3) {\color{black} b}
%	 \put (73,3) {\color{black} c}
%	 \end{overpic}}
%	\hfill
%	{\includegraphics[width=0.33\textwidth]{images/velocity_conservation2.jpg}}
%	\hfill
%	{\includegraphics[width=0.33\textwidth]{images/velocity_conservation3.jpg}}
%	\hfill
%	\caption{We initialized a volume with tangential surface velocity and advect the density field (color coded) and the velocity field. The volume on the left (a) has no vector correction, volume in the middle (b) with length correction from Eq.~\ref{eq:abs_momentum_conservation}. The static volume on the right (c) is for reference.}
%	\label{fig:vector_conservation}
%\end{figure*}

We refer the reader to the accompanying video to watch some of the examples in motion. Supplementary material discusses the vector length conservation experiment.
Moreover, we provide a precompiled version of our plugin for Houdini with two sample scenes for testing on Zenodo~\cite{Cappucino:2020:Samples}.

    \subsection{Performance}

	\begin{figure}
	    \centering
		\includegraphics[width=0.65\linewidth]{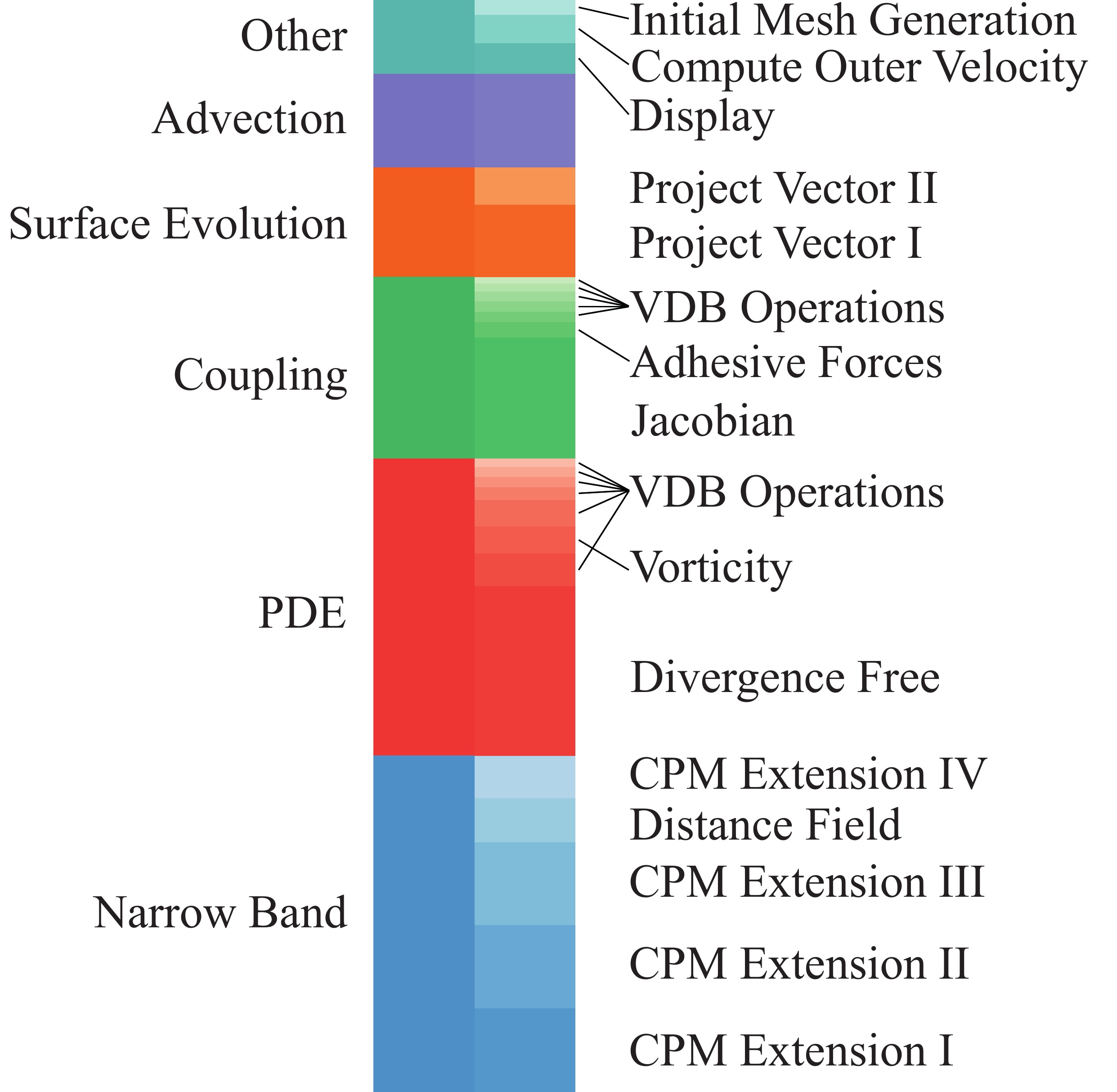}
		\caption{The calculation costs of our nodes for the CPM methods are of the same magnitude as the existing Houdini operators typically used in simulations.}
		\label{fig:performance}
	\end{figure}
    
     The performance was measured for every operator individually with the rotating sphere example (Fig.~\ref{fig:rotating_sphere}). 
     The grid resolution has the most significant impact on the computation time, but also the width of the narrow-band profoundly influences the performance. For the rotating sphere example, the overall computation time per frame with 5 substeps was $28\,$s. The soap film example from Fig.~\ref{fig:coffee} took approximately $28\,$s per frame with 2 substeps. The measurements were taken on a notebook with an Intel(R) Core i7-6700K CPU and a NVidia GeForce GTX 1070. The simulation of the underlying fluid is not included in the measurements. As can be seen in Fig.~\ref{fig:performance}, most of the computation time for a single task is spent in the \enquote{Divergence Free} operator. 
     This node uses the OpenVDBs Poisson solver with a maximum of 100 iterations to obtain a divergence-free vector field. Second most computation time is spent to compute the Jacobian for the coupling step. In third place is Houdini's advection operator. All other operators have smaller computation costs. This shows that the computation time for our steps is of the magnitude as Houdini's advection node.

	\section{Conclusion and Future Work}
	We presented a method to solve PDEs on evolving surfaces where we considered the external movement as the driving process. We derived physically motivated conservation laws and coupling strategies. We coupled simulations of 3D and 2D space in a way that allows us to compute high-resolution 2D simulations on coarse input surfaces efficiently and that exposes physically plausible parameters to control the strength of the coupling. We demonstrated that our approach can be used for different types of problems that require solving PDEs on a surface. We showed how to integrate all necessary steps into a VFX production environment in a modular way so that the building blocks can be reused and we provide a reference implementation as open source. The 2D surface simulation showed to be a usable tool for VFX creation that can add a big visual difference to scenes. 
	By using the CPM there are some limitations to our approach. As most CPM implementations, our method cannot model surfaces with hard edges or high frequencies. These limitations restrict use-cases to smooth surfaces. So far, we assume a surface without boundaries. Future research could integrate different boundary conditions. Testing further types of PDEs to achieve effects like droplet formation or fingering as presented by Vantzos et al.~\cite{azencot2015functional} but on moving surfaces is another topic for future work. Another area of interest is reinjecting 3D fluids from the surface as a simple form of two-way coupling of the simulations, e.g., for water droplets.  
	
	\section*{Acknowledgements}
	We want to thank Mariusz Wesierski for his help in setting up the river simulation and his support in all Houdini questions. This work is partly supported by \enquote{Kooperatives Promotionskolleg Digital Media} at Stuttgart Media University and the University of Stuttgart.
    %-------------------------------------------------------------------------
    \appendix
	\section{\textbf{Conservation of Intensive Scalar Quantities}}\label{sec:app1}
    
    In Section~\ref{susec:surf_evolve}, we stated that Eq.~\ref{eq:mass_cons_result} directly follows from Eq.~\ref{eq:1} if $a$ represents an intensive scalar quantity. 
    In the following, we provide a detailed derivation of this fact.
    We want to ensure that the total amount of $a$  does not change if there are no phenomena like mass transfer or chemical reactions present.
    As the surface evolves, the surface density $a$ needs to be adjusted to ensure that the total amount stays constant.
	Changing the differentiation and integration order in Eq.~\ref{eq:1} results in
	\begin{align}\label{eq:1_1}
	0 =\int_{M}\frac{D_O}{Dt}(a\,dM) =\int_{M}\left(\frac{D_O\,a}{Dt}dM + a\frac{D_O\,dM}{Dt}\right)\,.
	\end{align}
	Since $dM$ evolves, we need to evaluate the term $\frac{D_O\,dM}{Dt}$.

    According to Stone~\cite{stone1990simple}, the material derivative of a surface element can be described by
    \begin{align}\label{eq:id0}
	\frac{D}{Dt}dM=(\nabla_{T_\mathbf{p}M}\cdot\tilde{\mathbf{u}}_n)dM\,, 
	\end{align}
	where $\nabla_{T_\mathbf{p}M} = (I - \mathbf{n} \mathbf{n}^T)\nabla$ is the gradient operator projected onto the tangential space $T_\mathbf{p}M$ at point $\mathbf{p}$, and $\tilde{\mathbf{u}}_n$ is the velocity evolving the surface.
	Without loss of generality, we assume  that the outer process is described by $\mathbf{u}$ instead of $\tilde{\mathbf{u}}_n$ and use $\frac{D\,dM}{Dt}$ instead of $\frac{D_O\,dM}{Dt}$ to derive the rate of change of $dM$.
    To describe Eq.~\ref{eq:id0} in more detail, we first show that the identity
	\begin{align}\label{eq:id1}
	\frac{D}{Dt}d\mathbf{M} = (\nabla\cdot\mathbf{u})d\mathbf{M} - (\nabla\mathbf{u})^T d\mathbf{M}
	\end{align}
	holds for an arbitrary oriented surface element $d\mathbf{M}=\mathbf{n}dM$~\cite{batchelor1967}.
	To this end, we first investigate how volume elements $dV$ and line elements $d\mathbf{l}$ are changed due to the base animation.
	We consider the elements to be that small, that they are subject only to pure straining motion and rigid rotations~\cite{batchelor1967}.
	
	The rate of change of the volume element $dV$ is described by
	\begin{align}\label{eq:id2}
	    \frac{D\,dV}{Dt} = \nabla\cdot\mathbf{u}\,dV\,.
	\end{align}
	
	We assume that the line element $d\mathbf{l}$ is linear (and, therefore, can be described by a vector) and stays approximately straight.
	Under these assumptions, its rate of change is simply the difference of the velocities at the two ends of the element and can be described by
	\begin{align}\label{eq:id3}
	    \frac{D\,d\mathbf{l}}{Dt} = \nabla \mathbf{u} \, d\mathbf{l}\,.
	\end{align}
	The volume element $dV$ can also be described by $dV = d\mathbf{M}\cdot d\mathbf{l}$. 
	Inserting $dV$ into Eq.~\ref{eq:id2} results in:
	\begin{align}\label{eq:id4_2}
	     (\nabla\cdot\mathbf{u}) \,d\mathbf{M}\cdot d\mathbf{l} &= \nabla\cdot\mathbf{u}\,dV = \frac{D(d\mathbf{M} \cdot d\mathbf{l})}{Dt}\nonumber\\
	    &=d\mathbf{M}\cdot\frac{D\,d\mathbf{l}}{Dt} + \frac{D\,d\mathbf{M}}{Dt}\cdot d\mathbf{l}\nonumber\\
	    &= d\mathbf{M}\cdot(\nabla \mathbf{u}\, d\mathbf{l}) + \frac{D\,d\mathbf{M}}{Dt}\cdot d\mathbf{l}\nonumber\\
        &= ((\nabla \mathbf{u})^T d\mathbf{M})\cdot\, d\mathbf{l} + \frac{D\,d\mathbf{M}}{Dt}\cdot d\mathbf{l}\,.
	\end{align}
	As Eq.~\ref{eq:id4_2} has to hold for arbitrary line elements $d\mathbf{l}$, we obtain Eq.~\ref{eq:id1}.
	To get the formulation in Eq.~\ref{eq:id0} we take the inner-product of $\mathbf{n}$ with Eq.~\ref{eq:id1} and use the identity $\mathbf{n}^T(\nabla\mathbf{u})^T \mathbf{n}=(\mathbf{n}\mathbf{n}^T\nabla)\cdot\mathbf{u}$:
    \begin{align}
    \frac{D}{Dt}\,dM 
    &= \frac{D}{Dt}\mathbf{n}^T\mathbf{n}\,dM=\mathbf{n}^T\frac{D}{Dt}d\mathbf{M}\nonumber\\
	&= \mathbf{n}^T(\nabla\cdot\mathbf{u})\,d\mathbf{M} - \mathbf{n}^T(\nabla\mathbf{u})^T\, d\mathbf{M}\nonumber\\
% 	&=\mathbf{n}^T\mathbf{n} (\nabla\cdot\mathbf{u})\,dM - \mathbf{n}^T(\nabla\mathbf{u})^T \mathbf{n}\,dM\nonumber\\
	&= (\nabla\cdot\mathbf{u})\,dM - (\mathbf{n}\mathbf{n}^T\nabla)\cdot\mathbf{u}\, dM\nonumber\\
	&= ((I - \mathbf{n}\mathbf{n}^T)\nabla)\cdot\mathbf{u}\,dM = (\nabla_{T_\mathbf{p}M} \cdot \mathbf{u})\,dM\,.\label{eq:id5}
	\end{align}
	Inserting Eq.~\ref{eq:id0} into Eq.~\ref{eq:1_1} results in
	\begin{align}
		0&=\int_{M(t)} \left(\frac{D_Oa}{Dt}dM +a(\nabla_{T_\mathbf{p}M}\cdot\tilde{\mathbf{u}}_\mathbf{n})dM\right)\nonumber\\
		&= \int_{M(t)} \left(\frac{D_Oa}{Dt} + a(\nabla_{T_\mathbf{p}M}\cdot\tilde{\mathbf{u}}_\mathbf{n})\right)dM.\label{eq:integral_formualtion_divergence}
	\end{align}
	As Eq.~\ref{eq:integral_formualtion_divergence} holds for an arbitrary surface $M$, we get
	\begin{align}
	0 &= \frac{D_Oa}{Dt} + a(\nabla_{T_\mathbf{p}M}\cdot\tilde{\mathbf{u}}_\mathbf{n}) =  \frac{D_Oa}{Dt} + a(\nabla\cdot(\tilde{\mathbf{u}}_\mathbf{n})_{T_\mathbf{p}M}).
	\end{align}
	Therefore, the rate of change of a intensive quantity $a$ is defined by
	\begin{align}
	\frac{D_Oa}{Dt} &= - a(\nabla\cdot(\tilde{\mathbf{u}}_\mathbf{n})_{T_\mathbf{p}M}).
	\end{align}
	
	\section{\textbf{Length-preserving Evolution of a Vector Field}}\label{sec:app_length_preserving}

	In Section~\ref{susec:surf_evolve}, we stated that Eq.~\ref{eq:abs_momentum_conservation} is satisfied when setting
	\begin{align}\label{eq:app_final_velocity}
        \frac{D_O\mathbf{v}}{Dt} = \nabla \tilde{\mathbf{u}}_\mathbf{n}\mathbf{v} - \mathbf{v}\cdot\nabla\tilde{\mathbf{u}}_\mathbf{n}\mathbf{v}\frac{\mathbf{v}}{\|\mathbf{v}\|^2}.
	\end{align}
	To confirm that Eq.~\ref{eq:app_final_velocity} results from Eq.~\ref{eq:abs_momentum_conservation},
	we first show that $\frac{D_O\|\mathbf{v}\|}{Dt}=0$ follows from Eq.~\ref{eq:abs_momentum_conservation}.
	To this end, we reformulate the right-hand side of Eq.~\ref{eq:abs_momentum_conservation} using Eq.~\ref{eq:mass_cons_result} and Eq.~\ref{eq:id0}:
	\begin{align}
		0 &=\frac{D_O}{Dt}\int_M\norm{a\mathbf{v}}dM = \frac{D_O}{Dt}\int_Ma\norm{\mathbf{v}}dM\nonumber\\
		&= \int_M\left(\frac{D_Oa}{Dt}\norm{\mathbf{v}}dM + a\frac{D_O\norm{\mathbf{v}}}{Dt}dM + a\norm{\mathbf{v}}\frac{D_OdM}{Dt}\right)\nonumber\\
		&=\int_M a\frac{D_O\norm{\mathbf{v}}}{Dt}dM.
	\end{align}
	This is true for an arbitrary surface $M$ and we obtain: 
	\begin{align}\label{eq:momentum_conservation_result_appendix}
		\frac{D_O\norm{\mathbf{v}}}{Dt} = 0.
	\end{align}
	
	To achieve this, we evolve $\mathbf{v}$ with $\nabla\tilde{\mathbf{u}}_\mathbf{n}$ and compensate length changes. 
	The change in length of $\mathbf{v}$ can be expressed by 
	\begin{align}
		\frac{\mathbf{v}\cdot\nabla\tilde{\mathbf{u}}_\mathbf{n}\mathbf{v}}{\|\mathbf{v}\|}\frac{\mathbf{v}}{\|\mathbf{v}\|} \,, 
	\end{align}
	and we obtain:
	\begin{align}
		\frac{D_O\mathbf{v}}{Dt} = \nabla\tilde{\mathbf{u}}_\mathbf{n}\mathbf{v} - \left(\mathbf{v}\cdot\nabla\tilde{\mathbf{u}}_\mathbf{n}\mathbf{v}\right)\frac{\mathbf{v}}{\|\mathbf{v}\|^2}.
	\end{align}
	This can be shown when writing the derivative as the limit of the difference quotients. 
	To this end, we use the first-order Taylor expansion of $\mathbf{v}$ with respect to $t$ to define $\hat{\mathbf{v}} = \mathbf{v} + \Delta t \nabla \tilde{\mathbf{u}}_\mathbf{n}\mathbf{v}$.
	Compensating for length changes, $\mathbf{v}' := \mathbf{v}(t+\Delta t)$ can be approximated by:
	\begin{align}
		\mathbf{v}' = \hat{\mathbf{v}} - \left(\|\hat{\mathbf{v}}\|-\|\mathbf{v}\|\right)\frac{\hat{\mathbf{v}}}{\|\hat{\mathbf{v}}\|}
	\end{align}
	and the derivative of $\mathbf{v}$ is written as:
	\begin{align}
		\frac{D_O\mathbf{v}}{Dt} =& \lim\limits_{\Delta t\rightarrow 0}\frac{\mathbf{v}' - \mathbf{v}}{\Delta t}
		= \lim\limits_{\Delta t\rightarrow 0}\frac{1}{\Delta t}\left(\Delta t \nabla \tilde{\mathbf{u}}_\mathbf{n}\mathbf{v}-  \left(\|\hat{\mathbf{v}}\|-\|\mathbf{v}\|\right)\frac{\hat{\mathbf{v}}}{\|\hat{\mathbf{v}}\|}\right)\nonumber\\
% 		=& \lim\limits_{\Delta t\rightarrow 0}\left(\nabla \tilde{\mathbf{u}}_\mathbf{n}\mathbf{v} - \frac{1}{\Delta t}\left(1 - \frac{\|\mathbf{v}\|}{\|\hat{\mathbf{v}}\|} \right)\hat{\mathbf{v}}\right)\nonumber\\
		=& \lim\limits_{\Delta t\rightarrow 0}\left(\nabla \tilde{\mathbf{u}}_\mathbf{n}\mathbf{v} - \frac{1}{\Delta t}\left(1 - \frac{\|\mathbf{v}\|}{\|\hat{\mathbf{v}}\|} \right)\mathbf{v}+\left(1 - \frac{\|\mathbf{v}\|}{\|\hat{\mathbf{v}}\|} \right) \nabla \tilde{\mathbf{u}}_\mathbf{n}\mathbf{v}\right)\nonumber\\
% 		=& \lim\limits_{\Delta t\rightarrow 0}\left(\nabla \tilde{\mathbf{u}}_\mathbf{n}\mathbf{v} - \frac{\|\hat{\mathbf{v}}\| - \|\mathbf{v}\|}{\Delta t \|\hat{\mathbf{v}}\|} \mathbf{v}+\right.\nonumber\\ 
% 		& \left.\left(1 - \frac{\|\mathbf{v}\|}{\|\hat{\mathbf{v}}\|} \right) \nabla \tilde{\mathbf{u}}_\mathbf{n}\mathbf{v}\right)\nonumber\\
		=& \lim\limits_{\Delta t\rightarrow 0}\left(\nabla \tilde{\mathbf{u}}_\mathbf{n}\mathbf{v} - \frac{\|\hat{\mathbf{v}}\|^2 - \|\mathbf{v}\|^2}{\Delta t \|\hat{\mathbf{v}}\|\left(\|\hat{\mathbf{v}}\| + \|\mathbf{v}\|\right)} \mathbf{v}+\right.\nonumber\\ 
		& \left.\left(1 - \frac{\|\mathbf{v}\|}{\|\hat{\mathbf{v}}\|} \right) \nabla \tilde{\mathbf{u}}_\mathbf{n}\mathbf{v}\right)\nonumber\\
		=& \lim\limits_{\Delta t\rightarrow 0}\left(\nabla \tilde{\mathbf{u}}_\mathbf{n}\mathbf{v} - \frac{2\Delta t \mathbf{v}\cdot(\nabla\tilde{\mathbf{u}}_\mathbf{n}\mathbf{v}) + \Delta t^2\|\nabla\tilde{\mathbf{u}}_\mathbf{n}\mathbf{v}\|^2 }{\Delta t \|\hat{\mathbf{v}}\|\left(\|\hat{\mathbf{v}}\| + \|\mathbf{v}\|\right)} \mathbf{v}+\right.\nonumber\\ 
		& \left.\left(1 - \frac{\|\mathbf{v}\|}{\|\hat{\mathbf{v}}\|} \right) \nabla \tilde{\mathbf{u}}_\mathbf{n}\mathbf{v}\right)\nonumber\\
		=& \lim\limits_{\Delta t\rightarrow 0}\left(\nabla \tilde{\mathbf{u}}_\mathbf{n}\mathbf{v} - \frac{2\mathbf{v}\cdot(\nabla\tilde{\mathbf{u}}_\mathbf{n}\mathbf{v}) + \Delta t\|\nabla\tilde{\mathbf{u}}_\mathbf{n}\mathbf{v}\|^2 }{\|\hat{\mathbf{v}}\|\left(\|\hat{\mathbf{v}}\| + \|\mathbf{v}\|\right)} \mathbf{v}+\right.\nonumber\\ 
		& \left.\left(1 - \frac{\|\mathbf{v}\|}{\|\hat{\mathbf{v}}\|} \right) \nabla \tilde{\mathbf{u}}_\mathbf{n}\mathbf{v}\right)
		= \nabla \tilde{\mathbf{u}}_\mathbf{n}\mathbf{v} - \frac{\mathbf{v}\cdot(\nabla\tilde{\mathbf{u}}_\mathbf{n}\mathbf{v})}{\|\mathbf{v}\|^2}\mathbf{v}
	\end{align}
	%-------------------------------------------------------------------------

 	%\bibliographystyle{eg-alpha}
	\bibliographystyle{eg-alpha-doi}
	
	\bibliography{gvdb-bibliography.bib}

\newcommand{\etalchar}[1]{$^{#1}$}
\begin{thebibliography}{\uppercase{MRWE20b}}

\bibitem[ADAT13]{akinci2013screen}
\textsc{Akinci N., Dippel A., Akinci G., Teschner M.}:
\newblock Screen space foam rendering.
\newblock \emph{Journal of WSCG 21}, 3 (2013), 173--182.

\bibitem[AMT{\etalchar{*}}12]{auer2012real}
\textsc{Auer S., Macdonald C., Treib M., Schneider J., Westermann R.}:
\newblock Real-time fluid effects on surfaces using the closest point method.
\newblock \emph{Computer Graphics Forum 31}, 6 (2012), 1909--1923.
\newblock \href {https://doi.org/10.1111/j.1467-8659.2012.03071.x}
  {\path{doi:10.1111/j.1467-8659.2012.03071.x}}.

\bibitem[AVW{\etalchar{*}}15]{azencot2015functional}
\textsc{Azencot O., Vantzos O., Wardetzky M., Rumpf M., Ben-Chen M.}:
\newblock Functional thin films on surfaces.
\newblock In \emph{Proceedings of the 14th ACM SIGGRAPH / Eurographics
  Symposium on Computer Animation} (2015), pp.~137--146.
\newblock \href {https://doi.org/10.1145/2786784.2786793}
  {\path{doi:10.1145/2786784.2786793}}.

\bibitem[AW13]{auer2013semi}
\textsc{Auer S., Westermann R.}:
\newblock A semi-{L}agrangian closest point method for deforming surfaces.
\newblock \emph{Computer Graphics Forum 32}, 7 (2013), 207--214.
\newblock \href {https://doi.org/10.1111/cgf.12228}
  {\path{doi:10.1111/cgf.12228}}.

\bibitem[AWO{\etalchar{*}}14]{azencot2014functional}
\textsc{Azencot O., Wei{\ss}mann S., Ovsjanikov M., Wardetzky M., Ben-Chen M.}:
\newblock Functional fluids on surfaces.
\newblock \emph{Computer Graphics Forum 33}, 5 (2014), 237--246.
\newblock \href {https://doi.org/10.1111/cgf.12449}
  {\path{doi:10.1111/cgf.12449}}.

\bibitem[Bat00]{batchelor1967}
\textsc{Batchelor G.~K.}:
\newblock \emph{An Introduction to Fluid Dynamics}, 8th~ed.
\newblock Cambridge Mathematical Library. Cambridge University Press, 2000.
\newblock \href {https://doi.org/10.1017/CBO9780511800955}
  {\path{doi:10.1017/CBO9780511800955}}.

\bibitem[BK15]{Bender:2015:DFSPH}
\textsc{Bender J., Koschier D.}:
\newblock Divergence-free smoothed particle hydrodynamics.
\newblock In \emph{Proceedings of the 14th ACM SIGGRAPH/Eurographics Symposium
  on Computer Animation} (2015), pp.~147--155.
\newblock \href {https://doi.org/10.1145/2786784.2786796}
  {\path{doi:10.1145/2786784.2786796}}.

\bibitem[Bou97]{boussinesq1897theorie}
\textsc{Boussinesq J.}:
\newblock \emph{Th{\'e}orie de l'{\'e}coulement tourbillonnant et tumultueux
  des liquides dans les lits rectilignes a grande section}, 1~ed.
\newblock Gauthier-Villars, 1897.

\bibitem[CPPK07]{cleary2007bubbling}
\textsc{Cleary P.~W., Pyo S.~H., Prakash M., Koo B.~K.}:
\newblock Bubbling and frothing liquids.
\newblock \emph{ACM Transactions on Graphics 26}, 3 (2007), 97:2--97:6.
\newblock \href {https://doi.org/10.1145/1276377.1276499}
  {\path{doi:10.1145/1276377.1276499}}.

\bibitem[FSJ01]{fedkiw:2001:vorticity}
\textsc{Fedkiw R., Stam J., Jensen H.~W.}:
\newblock Visual simulation of smoke.
\newblock In \emph{Proceedings of the 28th Annual Conference on Computer
  Graphics and Interactive Techniques} (2001), pp.~15--22.
\newblock \href {https://doi.org/10.1145/383259.383260}
  {\path{doi:10.1145/383259.383260}}.

\bibitem[GDP16]{gagnon2016dynamic}
\textsc{Gagnon J., Dagenais F., Paquette E.}:
\newblock Dynamic lapped texture for fluid simulations.
\newblock \emph{The Visual Computer 32}, 6 (2016), 901--909.
\newblock \href {https://doi.org/10.1007/s00371-016-1262-8}
  {\path{doi:10.1007/s00371-016-1262-8}}.

\bibitem[GS84]{GrayScott}
\textsc{Gray P., Scott S.}:
\newblock Autocatalytic reactions in the isothermal, continuous stirred tank
  reactor: Oscillations and instabilities in the system a + 2b $\rightarrow$
  3b; b $\rightarrow$ c.
\newblock \emph{Chemical Engineering Science 39}, 6 (1984), 1087--1097.
\newblock \href {https://doi.org/10.1016/0009-2509(84)87017-7}
  {\path{doi:10.1016/0009-2509(84)87017-7}}.

\bibitem[HH16]{hill2016efficient}
\textsc{Hill D.~J., Henderson R.~D.}:
\newblock Efficient fluid simulation on the surface of a sphere.
\newblock \emph{ACM Transactions on Graphics 35}, 2 (2016), 16:1--16:9.
\newblock \href {https://doi.org/10.1145/2879177} {\path{doi:10.1145/2879177}}.

\bibitem[HIK{\etalchar{*}}20]{HuangSoapBubblesSIGGRAPH2020}
\textsc{Huang W., Iseringhausen J., Kneiphof T., Qu Z., Jiang C., Hullin
  M.~B.}:
\newblock Chemomechanical simulation of soap film flow on spherical bubbles.
\newblock \emph{ACM Transactions on Graphics 39}, 4 (2020).
\newblock \href {https://doi.org/10.1145/3386569.3392094}
  {\path{doi:10.1145/3386569.3392094}}.

\bibitem[IAAT12]{ihmsen2012unified}
\textsc{Ihmsen M., Akinci N., Akinci G., Teschner M.}:
\newblock Unified spray, foam and air bubbles for particle-based fluids.
\newblock \emph{The Visual Computer 28}, 6 (2012), 669--677.
\newblock \href {https://doi.org/10.1007/s00371-012-0697-9}
  {\path{doi:10.1007/s00371-012-0697-9}}.

\bibitem[IBAT11]{ihmsen2011animation}
\textsc{Ihmsen M., Bader J., Akinci G., Teschner M.}:
\newblock Animation of air bubbles with {SPH}.
\newblock In \emph{Proceedings of the International Conference on Computer
  Graphics Theory and Applications} (2011), pp.~225--234.
\newblock \href {https://doi.org/10.5220/0003322902250234}
  {\path{doi:10.5220/0003322902250234}}.

\bibitem[ISN{\etalchar{*}}20]{isnhw2020soapfilm_with_thickness}
\textsc{Ishida S., Synak P., Narita F., Hachisuka T., Wojtan C.}:
\newblock A model for soap film dynamics with evolving thickness.
\newblock \emph{ACM Transactions on Graphics 39}, 4 (2020), 31:1--31:11.
\newblock \href {https://doi.org/10.1145/3386569.3392405}
  {\path{doi:10.1145/3386569.3392405}}.

\bibitem[KBST19]{Koschier:2019:SPHTutorial}
\textsc{Koschier D., Bender J., Solenthaler B., Teschner M.}:
\newblock Smoothed particle hydrodynamics techniques for the physics based
  simulation of fluids and solids.
\newblock In \emph{Eurographics 2019 - Tutorials} (2019), The Eurographics
  Association.
\newblock \href {https://doi.org/10.2312/egt.20191035}
  {\path{doi:10.2312/egt.20191035}}.

\bibitem[KDBB17]{Koschier2017}
\textsc{{Koschier} D., {Deul} C., {Brand} M., {Bender} J.}:
\newblock An hp-adaptive discretization algorithm for signed distance field
  generation.
\newblock \emph{IEEE Transactions on Visualization and Computer Graphics 23},
  10 (2017), 2208--2221.
\newblock \href {https://doi.org/10.1109/TVCG.2017.2730202}
  {\path{doi:10.1109/TVCG.2017.2730202}}.

\bibitem[KLKK12]{kim2012controlling}
\textsc{Kim P.-R., Lee H.-Y., Kim J.-H., Kim C.-H.}:
\newblock Controlling shapes of air bubbles in a multi-phase fluid simulation.
\newblock \emph{The Visual Computer 28}, 6 (2012), 597--602.
\newblock \href {https://doi.org/10.1007/s00371-012-0696-x}
  {\path{doi:10.1007/s00371-012-0696-x}}.

\bibitem[KTT13]{kim2013closest}
\textsc{Kim T., Tessendorf J., Thuerey N.}:
\newblock Closest point turbulence for liquid surfaces.
\newblock \emph{ACM Transactions on Graphics 32}, 2 (2013), 15:1--15:13.
\newblock \href {https://doi.org/10.1145/2451236.2451241}
  {\path{doi:10.1145/2451236.2451241}}.

\bibitem[MBT{\etalchar{*}}15]{mercier2015surface}
\textsc{Mercier O., Beauchemin C., Thuerey N., Kim T., Nowrouzezahrai D.}:
\newblock Surface turbulence for particle-based liquid simulations.
\newblock \emph{ACM Transactions on Graphics 34}, 6 (2015), 202:1--202:10.
\newblock \href {https://doi.org/10.1145/2816795.2818115}
  {\path{doi:10.1145/2816795.2818115}}.

\bibitem[MLJ{\etalchar{*}}13]{Museth:2013:OOD:2504435.2504454}
\textsc{Museth K., Lait J., Johanson J., Budsberg J., Henderson R., Alden M.,
  Cucka P., Hill D., Pearce A.}:
\newblock {OpenVDB}: An open-source data structure and toolkit for
  high-resolution volumes.
\newblock In \emph{ACM SIGGRAPH 2013 Courses} (2013), pp.~19:1--19:1.
\newblock \href {https://doi.org/10.1145/2504435.2504454}
  {\path{doi:10.1145/2504435.2504454}}.

\bibitem[MMR13]{macdonald2013simple}
\textsc{Macdonald C.~B., Merriman B., Ruuth S.~J.}:
\newblock Simple computation of reaction--diffusion processes on point clouds.
\newblock \emph{Proceedings of the National Academy of Sciences 110}, 23
  (2013), 9209--9214.
\newblock \href {https://doi.org/10.1073/pnas.1221408110}
  {\path{doi:10.1073/pnas.1221408110}}.

\bibitem[MRWE20a]{Cappucino:2020:Samples}
\textsc{Morgenroth D., Reinhardt S., Weiskopf D., Eberhardt B.}:
\newblock Application and sample scenes for efficient {2D} simulation on moving
  {3D} surfaces, 2020.
\newblock \href {https://doi.org/10.5281/zenodo.4009208}
  {\path{doi:10.5281/zenodo.4009208}}.

\bibitem[MRWE20b]{Cappucino:2020}
\textsc{Morgenroth D., Reinhardt S., Weiskopf D., Eberhardt B.}:
\newblock Source code for efficient {2D} simulation on moving {3D} surfaces.
\newblock \url{https://github.com/dimo3d/Cappucino}, 2020.

\bibitem[Red70]{Redlich:1970:IntensiveExtensive}
\textsc{Redlich O.}:
\newblock Intensive and extensive properties.
\newblock \emph{Journal of Chemical Education 47}, 2 (1970), 154--156.
\newblock \href {https://doi.org/10.1021/ed047p154.2}
  {\path{doi:10.1021/ed047p154.2}}.

\bibitem[RKEW19]{Reinhardt:2019:Consistent}
\textsc{Reinhardt S., Krake T., Eberhardt B., Weiskopf D.}:
\newblock {Consistent Shepard interpolation for SPH-based fluid animation}.
\newblock \emph{ACM Transaction on Graphics 38}, 6 (2019), 189:1--189:11.
\newblock \href {https://doi.org/10.1145/3355089.3356503}
  {\path{doi:10.1145/3355089.3356503}}.

\bibitem[RM08]{ruuth2008simple}
\textsc{Ruuth S.~J., Merriman B.}:
\newblock A simple embedding method for solving partial differential equations
  on surfaces.
\newblock \emph{Journal of Computational Physics 227}, 3 (2008), 1943--1961.
\newblock \href {https://doi.org/10.1016/j.jcp.2007.10.009}
  {\path{doi:10.1016/j.jcp.2007.10.009}}.

\bibitem[SABK08]{seychelles2008thermal}
\textsc{Seychelles F., Amarouchene Y., Bessafi M., Kellay H.}:
\newblock Thermal convection and emergence of isolated vortices in soap
  bubbles.
\newblock \emph{{Physical Review Letters} 100}, 14 (2008), 144501.
\newblock \href {https://doi.org/10.1103/PhysRevLett.100.144501}
  {\path{doi:10.1103/PhysRevLett.100.144501}}.

\bibitem[SGGM06]{Sud2006}
\textsc{Sud A., Govindaraju N., Gayle R., Manocha D.}:
\newblock Interactive {3D} distance field computation using linear
  factorization.
\newblock In \emph{Proceedings of the 2006 Symposium on Interactive 3D Graphics
  and Games} (2006), pp.~117--124.
\newblock \href {https://doi.org/10.1145/1111411.1111432}
  {\path{doi:10.1145/1111411.1111432}}.

\bibitem[Sta03]{stam2003flows}
\textsc{Stam J.}:
\newblock Flows on surfaces of arbitrary topology.
\newblock \emph{ACM Transactions on Graphics 22}, 3 (2003), 724--731.
\newblock \href {https://doi.org/10.1145/1201775.882338}
  {\path{doi:10.1145/1201775.882338}}.

\bibitem[Sto90]{stone1990simple}
\textsc{Stone H.}:
\newblock A simple derivation of the time-dependent convective-diffusion
  equation for surfactant transport along a deforming interface.
\newblock \emph{Physics of Fluids A: Fluid Dynamics 2}, 1 (1990), 111--112.
\newblock \href {https://doi.org/10.1063/1.857686}
  {\path{doi:10.1063/1.857686}}.

\bibitem[SY04]{shi2004inviscid}
\textsc{Shi L., Yu Y.}:
\newblock Inviscid and incompressible fluid simulation on triangle meshes.
\newblock \emph{Computer Animation and Virtual Worlds 15}, 3-4 (2004),
  173--181.
\newblock \href {https://doi.org/10.1002/cav.19} {\path{doi:10.1002/cav.19}}.

\bibitem[TFK{\etalchar{*}}03]{takahashi2003realistic}
\textsc{Takahashi T., Fujii H., Kunimatsu A., Hiwada K., Saito T., Tanaka K.,
  Ueki H.}:
\newblock Realistic animation of fluid with splash and foam.
\newblock \emph{Computer Graphics Forum 22}, 3 (2003), 391--400.
\newblock \href {https://doi.org/10.1111/1467-8659.00686}
  {\path{doi:10.1111/1467-8659.00686}}.

\bibitem[TSS{\etalchar{*}}07]{thuerey2007real}
\textsc{Thuerey N., Sadlo F., Schirm S., M{\"u}ller-Fischer M., Gross M.}:
\newblock Real-time simulations of bubbles and foam within a shallow water
  framework.
\newblock In \emph{Proceedings of the 2007 ACM SIGGRAPH/Eurographics Symposium
  on Computer Animation} (2007), pp.~191--198.
\newblock \href {https://doi.org/10.2312/SCA/SCA07/191-198}
  {\path{doi:10.2312/SCA/SCA07/191-198}}.

\bibitem[XB14]{xu2014signed}
\textsc{Xu H., Barbi\v{c} J.}:
\newblock Signed distance fields for polygon soup meshes.
\newblock In \emph{Proceedings of Graphics Interface} (2014), pp.~35--41.

\bibitem[XZ03]{xu2003eulerian}
\textsc{Xu J.-J., Zhao H.-K.}:
\newblock {An Eulerian formulation for solving partial differential equations
  along a moving interface}.
\newblock \emph{Journal of Scientific Computing 19}, 1 (2003), 573--594.
\newblock \href {https://doi.org/10.1023/A:1025336916176}
  {\path{doi:10.1023/A:1025336916176}}.

\bibitem[ZB05]{ZhuBridson2005}
\textsc{Zhu Y., Bridson R.}:
\newblock Animating sand as a fluid.
\newblock \emph{ACM Transactions on Graphics 24}, 3 (2005), 965--972.
\newblock \href {https://doi.org/10.1145/1073204.1073298}
  {\path{doi:10.1145/1073204.1073298}}.

\end{thebibliography}
    %\printbibliography   
	%-------------------------------------------------------------------------

\end{document}